\documentclass[prd,aps,twocolumn,a4paper,showkeys,nofootinbib,10pt]{revtex4-1}

\usepackage{amsmath}
\usepackage{amsfonts}
\usepackage{amssymb}	
\usepackage{graphicx}
\usepackage{bm}
\usepackage{color}
\usepackage{commath}
\usepackage{hyperref}
\usepackage[T1]{fontenc}
\usepackage{xcolor}
\usepackage{verbatim}
\usepackage{comment}
\usepackage{float}
\usepackage{amsbsy}
\usepackage{makecell}
\usepackage{multirow}

\usepackage[normalem]{ulem}
\allowdisplaybreaks

\newcommand{\ls}[1]{{\textcolor{magenta}{\texttt{LS: #1}} }}

\newcommand{\tbd}[1]{%
  \ifmmode
    \textcolor{lightgray}{\text{\texttt{#1}}}%
  \else
    \textcolor{lightgray}{\texttt{#1}}%
  \fi
}

\newcommand{\orcid}[1]{\href{https://orcid.org/#1}{
\includegraphics[width=10pt]{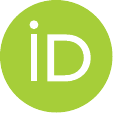}
}}

\begin{document}
\title{Exploring the gauge flexibility of the linear-in-spin effective-one-body Hamiltonian at the 5.5 post-Newtonian order}

\author{Andrea Placidi$\,{}^{1}$}
\author{Luca Sebastiani$\,{}^{2}$}
\author{Gianluca Grignani$\,{}^{1}$}

\affiliation{${}^{1}$Dipartimento di Fisica e Geologia, Universit\`a di Perugia,
\\
I.N.F.N. Sezione di Perugia, Via Pascoli, I-06123 Perugia, Italy}
\affiliation{${}^{2}$Max Planck Institute for Gravitational Physics (Albert Einstein Institute),
\\
Am M\"uhlenberg 1, Potsdam 14476, Germany}

\email{andrea.placidi7@gmail.com}
\email{luca.sebastiani@aei.mpg.de}
\email{gianluca.grignani@unipg.it}

\begin{abstract}

We derive the gauge-general expressions of the two gyro-gravitomagnetic functions entering the spin–orbit sector of the effective-one-body (EOB) Hamiltonian up to the fifth-and-half post-Newtonian (5.5PN) order. Our results include both local and nonlocal-in-time contributions, providing the most general analytical formulation of the linear-in-spin conservative dynamics within the EOB framework. These expressions are then employed to compute two gauge-invariant observables for quasi-circular orbits: the binding energy and the fractional periastron advance. We also use them to compare two spin gauge choices: the well-known Damour–Jaranowski–Schäfer (DJS) gauge, in which the gyro-gravitomagnetic functions are independent of the orbital angular momentum, and the alternative anti-DJS (or $\overline{\rm DJS}$) gauge, designed to reproduce in the test-mass limit the spin–orbit interaction of a spinning test particle in a Kerr background. For a circular, equal-mass, equal-spin binary, our analysis indicates that the $\overline{\rm DJS}$ gauge provides a slightly improved description of the inspiral dynamics, suggesting potential advantages for its use in future EOB waveform models.
\end{abstract}
\maketitle
\newpage
\section{Introduction}
\label{sec:intro}



The detection and detailed characterization of gravitational-wave (GW) signals from the coalescence of compact binaries are essential for inferring the properties of these systems, understanding their formation channels, and performing precise tests of general relativity (GR) \cite{LIGOScientific:2018mvr,LIGOScientific:2020ibl,LIGOScientific:2021djp, Nitz:2021zwj, Olsen:2022pin}. Achieving these goals requires continually improving analytical waveform models. A key element of this modeling effort is the accurate description of spin effects in the orbital dynamics of compact binaries \cite{Tagoshi:2000zg,Porto:2005ac,Faye:2006gx,Blanchet:2006gy,Porto:2006bt,Damour:2007nc,Hergt:2008jn,Porto:2008jj,Porto:2008tb,Perrodin:2010dy,Porto:2010tr,Hergt:2010pa,Blanchet:2011zv,Hartung:2011te,Hartung:2011ea,Levi:2011eq,Marsat:2012fn,Bohe:2012mr,Hartung:2013dza,Levi:2014gsa,Levi:2015uxa,Levi:2015ixa,Levi:2016ofk,Vines:2016qwa,Levi:2019kgk,Siemonsen:2019dsu,Levi:2020kvb,Levi:2020lfn,Cho:2021mqw, Bautista:2024agp}, as these effects leave a significant and observable imprint on the emitted gravitational radiation.

Several analytical approximation schemes have been developed to describe the conservative dynamics of compact binaries in GR. The post-Newtonian (PN) formalism~\cite{Blanchet:2013haa,Levi:2015msa,Porto:2016pyg,Levi:2018nxp,Schafer:2018kuf} is an expansion valid in the slow-motion, large-separation (i.e., weak-field) regime and is typically organized as a series in inverse powers of the speed of light, where a term proportional to $1/c^{n}$ corresponds to the $\tfrac{n}{2}$PN order. The post-Minkowskian (PM) framework~\cite{Westpfahl:1979gu,Bel:1981be,Damour:2016gwp,Damour:2017zjx}, by contrast, expands in the gravitational constant $G$ and applies at arbitrary velocities while still requiring sufficiently large separations. Finally, the gravitational self-force (GSF) approach~\cite{Mino:1996nk,Quinn:1996am,Barack:2018yvs,Pound:2021qin} relies on a perturbative expansion in the small mass ratio and is therefore restricted in mass range, while remaining valid in both strong-field and high-velocity regimes.

Building on one or more of these approximation schemes, the effective one-body (EOB) formalism~\cite{Buonanno:2000ef,Damour:2000we,Damour:2001tu,Damour:2015isa} provides a powerful framework that (i) anchors the description of binary evolution to the strong-field dynamics of a test body, thereby extending the domain of validity of the underlying perturbative information, and (ii) consistently incorporates calibrations to numerical relativity (NR) simulations. The result is a semi-analytical approach capable of generating accurate inspiral–merger–ringdown waveforms for coalescing binaries of any type, including systems with spin, tidal interactions, and eccentric orbits~\cite{Akcay:2020qrj,Schmidt:2020yuu,Nagar:2020xsk,Ossokine:2020kjp,Godzieba:2020bbz,Gamba:2020ljo,Gamba:2020wgg,Matas:2020wab,Riemenschneider:2021ppj,Gamba:2021ydi,Gamba:2020ljo,Gamba:2021gap,Ramos-Buades:2021adz,Bonino:2022hkj,Placidi:2021rkh,Albanesi:2022xge,Gonzalez:2022prs,Placidi:2023ofj,Albanesi:2023bgi,Grilli:2024lfh,Nagar:2024oyk,Damour:2025uka,Albanesi:2025txj,Nagni:2025cdw,pompili,seobnrv5hm,seobnrv5ehm_1,seobnrv5ehm_2,seobnrv5phm}.

In the conservative sector, the EOB dynamics is typically encoded in a Hamiltonian
\begin{equation} \label{eq:H_EOB}
H_{\rm EOB}=M\sqrt{1+2\nu(\hat{H}{\rm eff}-1)} ,
\end{equation}
where the rescaled effective Hamiltonian $\hat{H}{\rm eff} \equiv H_{\rm eff}/\mu$ governs the motion in the effective problem associated with a binary system of total mass $M=m_1+m_2$, reduced mass $\mu = m_1 m_2/M$, and symmetric mass ratio $\nu = \mu/M$. Neglecting contributions beyond linear order in the spins and considering spin-aligned (or anti-aligned) binaries, $\hat{H}_{\rm eff}$ can be decomposed as \cite{Damour:2008qf}
\begin{align}
\label{eq:Heff_general}
	&\hat{H}_{\rm eff} =\hat{H}_{\rm eff}^{\rm orb}(R,P_\varphi,P_R) \cr&\quad+ {P}_\varphi \left[ G_S(R,P_\varphi,P_R) \, S +  G_{S_*}(R,P_\varphi,P_R) \, {S}_* \right],
\end{align}
where $R$ is the relative separation between the two bodies, $P_\varphi$ is the orbital angular momentum, $P_R$ is the radial momentum, and $(S, S_*)$ denote the magnitudes of the linear combinations of the individual spins $({\bf S}_1,{\bf S}_2)$ defined below in Eq.~\eqref{eq:S_and_Sstar}. In the decomposition \eqref{eq:Heff_general}, the orbital Hamiltonian $\hat{H}_{\rm eff}^{\rm orb}$, describing the non-spinning component of the conservative dynamics, is augmented by a spin–orbit term (second line) whose coupling strength is encoded in the two effective gyro-gravitomagnetic functions $G_S$ and $G_{S_*}$.

Within the EOB framework, the functions $G_S$ and $G_{S_*}$ are fixed by matching the linear-in-spin results for the conservative two-body dynamics, as derived in any of the standard perturbative approaches. For instance, Refs.~\cite{Buonanno:2024byg,Damour:2025uka} determine them using PM-gravity results. In this work, instead, we focus on their determination within PN theory, whose linear-in-spin sector has been systematically developed across numerous studies.

In particular, Ref.~\cite{Damour:2008qf} derived $G_S$ and $G_{S_*}$ at next-to-leading order, corresponding to 2.5PN accuracy in the Hamiltonian, a result later extended to 3.5PN in Refs.~\cite{Nagar:2011fx,Barausse:2011ys}. By adapting the “Tutti Frutti’’ method~\cite{Bini:2019nra,Bini:2020wpo,Bini:2020nsb,Bini:2020hmy} to spinning systems, Refs.~\cite{Antonelli:2020aeb,Antonelli:2020ybz} further determined $G_S$ and $G_{S_*}$ at 4.5PN order. More recently, the same strategy was employed in Ref.~\cite{Khalil:2021fpm} to push the accuracy to 5.5PN.

It is important to recall that the explicit form of $G_S$ and $G_{S_*}$ is, in general, influenced by arbitrary choices of spin gauge. In Ref.~\cite{Damour:2008qf}, this freedom was used to impose the simplifying condition that both functions be independent of the angular momentum $P_\varphi$, thereby defining what is now known as the Damour–Jaranowski–Schäfer (DJS) gauge. While Refs.~\cite{Nagar:2011fx,Barausse:2011ys} computed $G_S$ and $G_{S_*}$ in a general, gauge-unfixed form, the higher-PN-order results of Refs.~\cite{Antonelli:2020aeb,Antonelli:2020ybz,Khalil:2021fpm} were obtained exclusively in the DJS gauge. This motivated the recent analysis of Ref.~\cite{Placidi:2024yld}, which derived the 4.5PN-accurate, gauge-unfixed expressions of $G_S$ and $G_{S_*}$ and carried out a first exploratory investigation of the potential advantages of adopting a spin gauge alternative to the DJS one, namely the so-called “anti-DJS’’ (or $\overline{\rm DJS}$) gauge.

Prompted also by Ref.~\cite{Albertini:2024rrs}, which highlighted the benefits of the $\overline{\rm DJS}$ gauge in the large–mass-ratio region of the parameter space, in the present work we extend the analysis of Ref.~\cite{Placidi:2024yld} by (i) pushing the gauge-general determination of the gyro-gravitomagnetic functions up to 5.5PN order, matching the accuracy of the DJS-gauge computation of Ref.~\cite{Khalil:2021fpm}, and (ii) providing the correspondingly accurate result within the $\overline{\rm DJS}$ gauge. We then assess the performance of the two spin gauges by comparing the associated linear-in-spin binding energy for circular orbits.

As already shown in Ref.~\cite{Khalil:2021fpm}, tail-transported nonlocal-in-time terms appear in the 5.5PN spin–orbit sector of the conservative dynamics, just as they do in the orbital sector at 4PN order. We therefore follow the standard strategy of treating the local-in-time and nonlocal-in-time effects separately. At the level of the gyro-gravitomagnetic functions, this corresponds to the 5.5PN decomposition
\begin{subequations}
\label{eq:gyrogfunctions_5.5PN}
\begin{align}
    & G_S^{\rm 5.5PN} = G_S^{\rm 5.5PN,loc}+G_S^{\rm 5.5PN, nonloc}, \label{full_gs} \\
    & G_{S_*}^{\rm 5.5PN} = G_{S_*}^{\rm 5.5PN,loc} +G_{S_*}^{\rm 5.5PN, nonloc} \label{full_gsstar},
\end{align}
\end{subequations}
where the local and nonlocal components, denoted respectively by the superscripts ``loc'' and ``nonloc'', are computed and provided separately.
 

Regarding the structure of the paper, we organize it as follows. In Sec.~\ref{sec:notation}, we clarify our notation and conventions. In Sec.~\ref{sec: non loc}, we address the nonlocal contributions arising from tail effects, implementing the Delaunay-averaging procedure at 5.5PN order. This is complemented in Sec.~\ref{sec: loc} by the derivation of the local-in-time component of the gyro-gravitomagnetic functions, which proceeds through the computation of the conservative, local-in-time part of the PN-expanded scattering angle.
Sec.~\ref{sec: gauge-i} is dedicated to the computation of the binding energy and periastron advance for quasi-circular orbits, which also serve as diagnostic tools to test the 5.5PN-accurate, gauge-general expressions for $G_S$ and $G_{S_*}$ obtained in the previous sections. These expressions are then employed in Sec.~\ref{sec:gauge_fixing} to derive the corresponding forms in the DJS and $\overline{\rm DJS}$ spin gauges, which we compare at the level of the spin–orbit contribution to the circular-orbit binding energy. Finally, we draw our conclusions in Sec.~\ref{sec:concl}.

We also supplement this paper with an Appendix, which reviews additional technical details relevant to the computation discussed in Sec.~\ref{sec: non loc}, and with a supplementary file~\cite{suppmat}, where we provide, in electronic form, all the main analytical results derived throughout this work.

\section{Computational setup}
\label{sec:notation}

We consider a system of two Kerr black holes with masses $m_1$ and $m_2$, total mass 
$M = m_1 + m_2$, and reduced mass $\mu = m_1 m_2/(m_1 + m_2)$.  
Let $\nu$ denote the symmetric mass ratio and $\delta$ the antisymmetric mass ratio, defined as
\begin{equation}
    \nu = \frac{\mu}{M}\,,
    \qquad
    \delta = \frac{m_1 - m_2}{M} = \sqrt{1 - 4\nu}\,,
\end{equation}
where, without loss of generality, we assume $m_1 \ge m_2$.

Let $\mathbf{S}_1$ and $\mathbf{S}_2$ denote the spins of the two black holes.  
In the EOB description of spinning binaries, the basic spin variables are: the Kerr parameters $\mathbf{a}_i = \mathbf{S}_i/m_i$; the dimensionless spins $\chi_i = |\mathbf{S}_i|/m_i^2$; the effective linear combinations
\begin{equation} \label{eq:S_and_Sstar}
    \mathbf{S} = \mathbf{S}_1 + \mathbf{S}_2\,,
    \qquad
    \mathbf{S}_* = \frac{m_2}{m_1}\,\mathbf{S}_1 + \frac{m_1}{m_2}\,\mathbf{S}_2\,;
\end{equation} the symmetric and antisymmetric dimensionless spin combinations
\begin{equation}
    \chi_S = \frac{\chi_1 + \chi_2}{2}\,,
    \qquad
    \chi_A = \frac{\chi_1 - \chi_2}{2}\,.
\end{equation}

When referring to the non-rescaled, dimensionful quantities, the relative separation, 
angular momentum, radial momentum, and Hamiltonian are denoted by 
$(R, P_R, P_\varphi, H)$. In rescaled, dimensionless form, these variables are
\begin{equation}
    r = \frac{R}{M}, \quad p_r = \frac{P_R}{\mu}, \quad p_\varphi = \frac{P_\varphi}{\mu M}, \quad \hat{H} = \frac{H}{\mu}.
    \label{scalign}
\end{equation}
The rescaled total momentum is, as usual, given by $p^2 = p_r^2 + p_\varphi^2 / r^2$.
It is also convenient to introduce the dimensionless variable $u \equiv 1/r$, which, in geometric units, coincides with the magnitude of the Newtonian potential.
Finally, for the gyro-gravitomagnetic functions we employ the $u$-rescaled combinations, $g_S \equiv G_S/u^3$ and $g_{S_*} \equiv G_{S_*}/u^3$.

In all our calculations, we restrict ourselves, for simplicity, to the case of 
spin-aligned (or anti-aligned) binaries. However, our results remain valid for 
generic precessing-spin configurations~\cite{spin_prec_Antonelli_2020}, since at 
linear order in the spins the gravitational spin-orbit coupling depends only on 
the scalar product between the spin and the orbital angular momentum, and is 
therefore insensitive to transverse spin components.

Because we focus exclusively on linear-in-spin effects, all higher-order 
spin contributions to the conservative dynamics are neglected throughout this work.

Finally, we adopt geometric units and set $G = c = 1$, restoring factors of 
$1/c$ only when needed to indicate the post-Newtonian order of 
PN-expanded expressions.

\section{Nonlocal spin–orbit dynamics in full gauge generality at 5.5PN}\label{sec: non loc}

In this section, we focus on the nonlocal 5.5PN components of the spin-orbit dynamics. We begin by recalling the procedure for computing the Delaunay-averaged Hamiltonian at 
5.5PN accuracy. We then show how to use it, combined with a suitable general ansatz in the nonlocal spin-orbit part of the effective Hamiltonian, to determine the gauge-general expressions of $g_S^{\rm 5.5PN, nonloc} = G_S^{\rm 5.5PN, nonloc}/u^3$ and $g_{S_*}^{\rm 5.5PN, nonloc} = G_{S_*}^{\rm 5.5PN, nonloc}/u^3$, namely the nonlocal components of the 5.5PN gyro-gravitomagnetic functions introduced in Eq.~\eqref{eq:gyrogfunctions_5.5PN}.

 
\subsection{Delaunay-averaged nonlocal Hamiltonian at 5.5PN}\label{subsec:harmonic}

For the sake of clarity and self-consistency, we begin by briefly reviewing the computation of the 5.5PN Delaunay-averaged Hamiltonian, originally carried out in Ref.~\cite{Khalil_2021} and independently reproduced by us.

It is well known that, starting at the fourth subleading post-Newtonian order 
(namely  at 4PN in the orbital sector and at 5.5PN in the spin-orbit sector), 
nonlocal-in-time contributions enter the conservative two-body dynamics. Accordingly, the action can be split into two components, $
\mathcal{S}_{\text{tot}}^{\text{nPN}} = \mathcal{S}_{\text{loc}}^{\text{nPN}} + \mathcal{S}_{\text{nonloc}}^{\text{nPN}}
$,
where the orbital part of $\mathcal{S}_{\text{nonloc}}^{\text{nPN}}$ has been computed up to $6$PN~\cite{4pn_eob_Damour_2015,5PN_Bini_2020,6PN_Bini_2020}.

The nonlocal part of the rescaled action, containing the leading-order contributions 
in both the orbital and spin-orbit sectors (at 4PN and 5.5PN, respectively), is given by
\begin{equation} \label{eq:Snonloc}
    \mathcal{S}_{\text{nonloc}}^{\text{LO}}
    = \int dt\, \text{Pf}_{2s/c} 
      \int \frac{dt'}{|t - t'|}\,
      \mathcal{F}_{\text{LO}}^{\text{sym}}(t, t')\,,
\end{equation}
where $\text{Pf}_{2s/c}$ denotes the Hadamard \textit{partie finie} 
regularization~\cite{hadamardBlanchet_2000} with arbitrary length scale $s$, 
introduced to regularize the singularity at $t' = t$, and 
$\mathcal{F}_{\text{LO}}^{\text{sym}}(t, t')$ is the time-symmetric gravitational-wave energy flux, restricted to its leading orbital 
and spin-orbit contributions.

Once the nonlocal action \eqref{eq:Snonloc} is evaluated, the corresponding
leading-order contribution to the Hamiltonian, denoted 
$\delta H_{\text{nonloc}}^{\text{LO}}$, follows directly from the
nonlocal term in the action and is related to it through
\begin{equation}
    \mathcal{S}_{\text{nonloc}}^{\text{LO}}
    = - \int dt\, \delta H_{\text{nonloc}}^{\text{LO}}(t)\,.
    \label{eq:nonloc}
\end{equation}

To evaluate Eq.~\eqref{eq:Snonloc}, we require the explicit form of the 
time-symmetric gravitational-wave energy flux,
\begin{equation}
    \mathcal{F}_{\text{LO}}^{\text{sym}}(t, t') =  \frac{1}{5} I_{ij}^{(3)}(t) I_{ij}^{(3)}(t') + \frac{16}{45c^2} J_{ij}^{(3)}(t) J_{ij}^{(3)}(t') ,
    \label{flussoLO}
\end{equation}
where $I_{ij}$ and $J_{ij}$ denote the mass- and current-type quadrupole moments.  
Retaining only the leading-order contributions that are spin-independent or linear 
in the spins, their expressions in the center-of-mass frame read 
\cite{Blanchetdamour,Kidder_1995}
\begin{subequations}
\begin{align}
    I_{ij} &= M \nu x^{\langle i} x^{j \rangle} 
+ \frac{1}{3 c^3} \left[ \frac{m_2^2}{M^2} \left( 4 v^{\langle i} ( \mathbf{S}_1 \times \bm{x} )^{j \rangle}\right.\right.
\cr&\left.\left.\hspace{0.4cm}- 5 x^{\langle i} ( \mathbf{S}_1 \times \bm{v} )^{j \rangle} \right) + 1 \leftrightarrow 2 \right], \label{mass_quad_cdm}\\
J_{ij} &= M \delta \nu x^{\langle i} ( \bm{x} \times \bm{v} )^{j \rangle} 
\cr &\hspace{0.4cm}+ \frac{3}{2c} \left[ \frac{m_2}{M} x^{\langle i} S_1^{j \rangle} - \frac{m_1}{M} x^{\langle i} S_2^{j \rangle} \right], \label{mass_curr_cdm}
\end{align}
\end{subequations}
where $\bm{x}$ and $\bm{v}$ denote the relative position and velocity of the 
binary, and angular brackets indicate the symmetric trace-free projection.

For the spin-aligned binaries we consider, the motion is planar. We can thus write
\begin{subequations}
\begin{align}
\bm{x} & = M\,r(\cos\varphi, \sin\varphi,0), \\
\bm{v} &= M\,\dot{r}(\cos\varphi, \sin\varphi,0) + Mr\dot{\varphi}(-\sin\varphi, \cos\varphi,0),
\end{align}
\end{subequations}
where we used mass-rescaled polar coordinates. It is then convenient to express $r$ and $\varphi$ in terms of orbital elements through the quasi-Keplerian parametrization, originally developed in Ref.~\cite{qk_DD_AIHPA_1985__43_1_107_0}, extended to 3PN accuracy in Ref.~\cite{qk2_Memmesheimer_2004}, and later generalized to include linear- and quadratic-in-spin contributions in Refs.~\cite{qkspin1_Tessmer_2010,qkspin2_Tessmer_2012}.
At $1$PN, the order we need in Eq.~\eqref{eq:Snonloc}, such parametrization is given by
\begin{subequations}
\label{eq:QKparam}
\begin{align}
r &= a_r (1 - e_r \cos u_e), \label{rqk} \\
\varphi &= 2K \arctan \left[ \sqrt{\frac{1 + e_\varphi}{1 - e_\varphi}} \tan \frac{u_e}{2} \right], \label{phi_qk}\\
\ell &= nt = u_e - e_t \sin u_e, \label{Kepler_quation}
\end{align}
\end{subequations}
where $a_r$ is the semi-major axis, $e_r$, $e_\varphi$, and $e_t$ denote respectively the 
radial, angular, and time eccentricities, $u_e$ is the 
eccentric anomaly, $n$ is the mean motion, and $K$ is the fractional periastron 
advance. For completeness, additional details on the quasi-Keplerian 
parametrization, as well as the relations between the orbital elements and the 
corresponding gauge-invariant quantities (at leading nonspinning and 
linear-in-spin order), are provided in Appendix~\ref{app.qk}.

When inserting the parametrization \eqref{eq:QKparam} into the quadrupole moments 
\eqref{mass_quad_cdm}–\eqref{mass_curr_cdm}, one must carefully handle the 
time derivatives. Up to relative 2.5PN order, the orbital elements 
$(a_r,\, e_r,\, e_\varphi,\, e_t,\, n,\, K)$ may be treated as constant in time, 
since each of them can be expressed in terms of energy and angular 
momentum [see Eqs.~\eqref{app:ar}–\eqref{app:ephi/et}], which are conserved quantities before radiation-reaction effects enter at 2.5PN order \cite{qk_DD_AIHPA_1985__43_1_107_0,qk_DD2}. 

Because the quasi-Keplerian parametrization is required here only up to 1.5PN 
accuracy, the eccentric anomaly $u_e$ is the only remaining time-dependent 
variable, and its evolution is governed by Kepler’s equation \eqref{Kepler_quation}, 
which yields
\begin{equation}
    \dot{u}_{\rm e} = \frac{n}{1 - e_t \cos u_e}\,.
\end{equation}
This relation allows one to straightforwardly order-reduce the time derivatives 
appearing in the quadrupole moments.

It is important to remark that the integral \eqref{eq:Snonloc} can only be 
analytically computed, for bound systems, within a small-eccentricity expansion. 
This is achieved by expressing the eccentricities $e_r$ and $e_\varphi$ in terms of 
$e_t$ (see App.~\ref{app.qk}) and then expanding in powers of $e_t$. In parallel, 
it is convenient to make the time dependence of $u_e$ explicit by solving 
Kepler’s equation in terms of Fourier--Bessel functions, namely
\begin{equation} \label{eq:Kepler}
    u_e = n t
    + \sum_{k=1}^{\infty} \frac{2}{k}\, J_k(k e_t)\, \sin(k n t)\,,
\end{equation}
where the Fourier series can be truncated consistently at the same order as the 
eccentricity expansion.



Once these operations are carried out on the energy flux in Eq.~\eqref{eq:Snonloc}, the Delaunay-averaged nonlocal Hamiltonian is obtained, as a 
function of $a_r$ and $e_t$, by averaging over one orbital period.\footnote{
This averaging procedure generates divergent integrals, which can be regularized, at this order, 
by taking their Hadamard \textit{partie finie}; see, e.g., Sec.~IV of 
Ref.~\cite{4pn_Damour_2014} for further details on this regularization scheme.}
In particular, by truncating the expansion in $e_t$ at eighth order, we exactly 
reproduce the Delaunay-averaged Hamiltonian 
$\left\langle \delta H_{\text{nonloc}}^{\text{LO}} \right\rangle$ 
given in Eq.~(2.19) of Ref.~\cite{Khalil_2021}.

\subsection{Nonlocal Delaunay-averaged EOB Hamiltonian at 5.5PN}\label{subsec:EOB}
The next step is to apply the Delaunay-averaging procedure at the level of the EOB 
Hamiltonian. Rewriting Eq.~\eqref{eq:H_EOB} in mass-rescaled form, we obtain
\begin{equation}
\hat{H}_{\text{EOB}} = \frac{1}{\nu} \sqrt{1 + 2\nu \left(\hat{H}_{\text{eff}} - 1\right)},
\label{energy_map_rescaled}
\end{equation}
where
\begin{align}
\hat{H}_{\text{eff}} = &\sqrt{A(u) \left[ 1 + p^2 + \left( A(u)\bar{D}(u) - 1 \right) p_r^2 + Q(u, p_r) \right]} \cr&
+ \frac{u^3 p_\varphi}{c^3}  \left[ g_S(u, p_r, p) \, S + g_{S_*}(r, p_r, p)\, S_* \right].
\label{H_eff_rescaled}
\end{align}

For our purposes, the EOB potentials $A(u)$, $\bar{D}(u)$, and $Q(u,p_{r})$ are 
taken at 4PN accuracy, including both their local and nonlocal parts. Among these, 
only the $Q$ potential requires an eccentricity truncation; we implement this by 
retaining terms up to eighth order in the eccentricity, which corresponds to 
keeping powers of $p_{r}$ up to $p_{r}^{8}$. For completeness, we report their 
explicit expressions below:
\begin{subequations}
\begin{align}
    A(u) &= 1 - 2u + a_3 u^3 + a_4 u^4 + (a_5^c + a_5^\text{log} \log u) u^5, \\
    \bar{D}(u) &= 1 + \bar{d}_2 u^2 + \bar{d}_3 u^3 + (\bar{d}_4^c + \bar{d}_4^\text{\,log} \log u) u^4, \\
    Q(u,p_r) &= q_{42}\,p_r^4 u^2 + q_{43}\,p_r^4 u^3 + q_{62}\,p_r^6 u^2 + q_{81}\,p_r^8 u,
\end{align}
\end{subequations}
where the coefficients read \cite{4pn_eob_Damour_2015,Bini:2020wpo}
\begin{subequations}
\begin{align}
a_3 &= 2\nu, \qquad 
a_4 = \left( \frac{94}{3} - \frac{41\pi^2}{32} \right) \nu, \qquad 
a_5^\text{log} = \frac{64}{5} \nu, \label{eob_coeff_primo} \\
a_5^c &= \left( \frac{2275\pi^2}{512} - \frac{4237}{60} + \frac{128}{5}\gamma_\mathrm{E} + \frac{256}{5} \ln 2 \right)\nu \cr 
&\hspace{0.4cm}
+ \left( \frac{41\pi^2}{32} - \frac{221}{6} \right)\nu^2, \\
\bar{d}_2 &= 6\nu, \qquad
\bar{d}_3 = 52\nu - 6\nu^2, \qquad
\bar{d}_4^\text{\,log} = \frac{592}{15} \nu, \\
\bar{d}_4^c &= \left( -\frac{533}{45} - \frac{23761\pi^2}{1536} + \frac{1184}{15} \gamma_\mathrm{E} 
- \frac{6496}{15} \ln 2\right.\cr 
&\hspace{0.4cm} \left.+ \frac{2916}{5} \ln 3 \right)\nu + \left( \frac{123\pi^2}{16} - 260 \right)\nu^2, \\
q_{42} &= 2(4 - 3\nu)\nu, \\
q_{43} &= \left( -\frac{5308}{15} + \frac{496\,256}{45} \ln 2 - \frac{33\,048}{5} \ln 3 \right) \nu \cr& - 83\nu^2 + 10\nu^3, \\
q_{62} &= \left( -\frac{827}{3} - \frac{2\,358\,912}{25} \ln 2 + \frac{1\,399\,437}{50} \ln 3\right.\cr &\hspace{0.4cm} \left. + \frac{390\,625}{18} \ln 5 \right) \nu - \frac{27}{5} \nu^2 + 6\nu^3, \\
q_{81} &= \left( -\frac{35\,772}{175} + \frac{21\,668\,992}{45} \ln 2 + \frac{6\,591\,861}{350} \ln 3\right.\cr &\hspace{0.4cm} \left. - \frac{27\,734\,375}{126} \ln 5 \right) \nu. \label{eob_coeff_ultimo}
\end{align}
\end{subequations}

To single out the nonlocal component of the effective Hamiltonian, each 
EOB potential, as well as the two gyro-gravitomagnetic functions, is split into a local 
and a nonlocal contribution, as in Eqs.~\eqref{eq:gyrogfunctions_5.5PN}.\footnote{This type of local/nonlocal decomposition was first 
introduced in Ref.~\cite{4pn_Damour_2014}.}
Treating the nonlocal parts as small corrections (entering at 4PN in the 
potentials and at 5.5PN in $g_S$ and $g_{S_*}$), we arrive at the following 
leading-order nonlocal contribution to the effective Hamiltonian:
\begin{align}
\label{H_EOB_nonloc}
&\delta \hat{H}_{\text{eff}}^{\text{nonloc}} = \frac{1}{2} \left( A^{\text{nonloc}} + \bar{D}^{\text{nonloc}}\, p_r^2 + Q^{\text{nonloc}} \right)
\cr &\hspace{0.2cm}+ \frac{p_\varphi u^3}{c^3 }  \left[ S\, g_S^{5.5\text{PN}, \text{nonloc}} + S_*\, g_{S_*}^{5.5\text{PN}, \text{nonloc}} \right].
\end{align}
Since our goal is to derive fully general expressions for $g_S$ and $g_{S_*}$, 
without imposing any \emph{a priori} spin–gauge choice, we begin by introducing a 
general ansatz for their nonlocal 5.5PN components, to be inserted into 
Eq.~\eqref{H_EOB_nonloc}. Focusing, for example, on $g_S$, the most general ansatz 
for its nonlocal part that is simultaneously compatible with the 5.5PN order and 
with the eighth order in eccentricity can be written as
\begin{widetext}
\begin{align}
g_S^{5.5\text{PN}, \text{nonloc}}&=\frac{1}{c^8}\bigg[\,g^{\text{N}^4\text{LO,nl}}_1\; u^4 
+ g^{\text{N}^4\text{LO,nl}}_2\; p_r^2 u^3 
+ g^{\text{N}^4\text{LO,nl}}_3\; \dot{p}_r^2 
+ g^{\text{N}^4\text{LO,nl}}_4\; p_r^4 u^2 
+ g^{\text{N}^4\text{LO,nl}}_5\; \frac{p_r^2 \dot{p}_r^2}{u} 
+ g^{\text{N}^4\text{LO,nl}}_6\; \frac{\dot{p}_r^4}{u^4} \cr
&+ g^{\text{N}^4\text{LO,nl}}_7\; p_r^6 u 
+ g^{\text{N}^4\text{LO,nl}}_8\; \frac{p_r^4 \dot{p}_r^2}{u^2} 
+ g^{\text{N}^4\text{LO,nl}}_9\; \frac{p_r^2 \dot{p}_r^4}{u^5} 
+ g^{\text{N}^4\text{LO,nl}}_{10}\; \frac{\dot{p}_r^6}{u^8} 
+ g^{\text{N}^4\text{LO,nl}}_{11}\; p_r^8 
+ g^{\text{N}^4\text{LO,nl}}_{12}\; \frac{p_r^6 \dot{p}_r^2}{u^3} \cr
&+ g^{\text{N}^4\text{LO,nl}}_{13}\; \frac{p_r^4 \dot{p}_r^4}{u^6} 
+ g^{\text{N}^4\text{LO,nl}}_{14}\; \frac{p_r^2 \dot{p}_r^6}{u^9} 
+ g^{\text{N}^4\text{LO,nl}}_{15}\; \frac{\dot{p}_r^8}{u^{12}} \cr
&+ \left( 
g^{\text{N}^4\text{LO,nl}}_{1,\,\text{log}}\; u^4 
+ g^{\text{N}^4\text{LO,nl}}_{2,\,\text{log}}\; p_r^2 u^3 
+ g^{\text{N}^4\text{LO,nl}}_{3,\,\text{log}}\; \dot{p}_r^2 
+ g^{\text{N}^4\text{LO,nl}}_{4,\,\text{log}}\; p_r^4 u^2 
+ g^{\text{N}^4\text{LO,nl}}_{5,\,\text{log}}\; \frac{p_r^2 \dot{p}_r^2}{u} 
+ g^{\text{N}^4\text{LO,nl}}_{6,\,\text{log}}\; \frac{\dot{p}_r^4}{u^4} \right. \cr
&\left. 
+ g^{\text{N}^4\text{LO,nl}}_{7,\,\text{log}}\; p_r^6 u 
+ g^{\text{N}^4\text{LO,nl}}_{8,\,\text{log}}\; \frac{p_r^4 \dot{p}_r^2}{u^2} 
+ g^{\text{N}^4\text{LO,nl}}_{9,\,\text{log}}\; \frac{p_r^2 \dot{p}_r^4}{u^5} 
+ g^{\text{N}^4\text{LO,nl}}_{10,\,\text{log}}\; \frac{\dot{p}_r^6}{u^8} 
+ g^{\text{N}^4\text{LO,nl}}_{11,\,\text{log}}\; p_r^8 
+ g^{\text{N}^4\text{LO,nl}}_{12,\,\text{log}}\; \frac{p_r^6 \dot{p}_r^2}{u^3} \right. \cr
&\left. 
+ g^{\text{N}^4\text{LO,nl}}_{13,\,\text{log}}\; \frac{p_r^4 \dot{p}_r^4}{u^6} 
+ g^{\text{N}^4\text{LO,nl}}_{14,\,\text{log}}\; \frac{p_r^2 \dot{p}_r^6}{u^9} 
+ g^{\text{N}^4\text{LO,nl}}_{15,\,\text{log}}\; \frac{\dot{p}_r^8}{u^{12}} 
\right) \log u\bigg],
\label{ansatz_gs}
\end{align}
\end{widetext}
where $(g_{n}^{\text{N}^4\text{LO,nl}},g_{n,\text{log} }^{\text{N}^4\text{LO,nl}})$ form a set of 30 $\nu$-dependent dimensionless coefficients. An analogous ansatz, involving 
another set of 30 coefficients $(g_{*n}^{\text{N}^4\text{LO,nl}},g_{*n,\text{log} }^{\text{N}^4\text{LO,nl}})$, is introduced for $g_{S_*}^{5.5\text{PN}, \text{nonloc}}$.

We remark that we have chosen the dynamical variables $(u,\, p_r,\, \dot{p}_r)$ in 
place of the more natural set $(u,\, p_r,\, p^{2})$, with 
$p^{2} = p_r^{2} + p_\varphi^{2} u^{2}$. This choice is motivated by the 
small-eccentricity expansion: our variables must also behave as consistent 
power counters in $e_t$. While $u = \mathcal{O}(e_t^{0})$ and 
$p_r = \mathcal{O}(e_t) = \dot{p}_r$, the eccentricity counting is less 
transparent when using $p$. Nevertheless, once the matching is performed, one 
may always convert $\dot{p}_r$ back to $p^{2}$ by inverting the EOB Hamilton 
equation for $p_r$ at the required PN accuracy.

Finally, the inclusion of logarithmic terms in the ansatz is necessary because, 
at the same PN order, the Delaunay-averaged nonlocal Hamiltonian obtained in the previous 
section contains contributions proportional to $\log a_r$.

After inserting the ansatz into $\delta \hat{H}_{\text{eff}}^{\text{nonloc}}$, 
the next step is to perform the Delaunay averaging. To this end, we use in 
Eq.~\eqref{H_EOB_nonloc} the Newtonian relations
\begin{equation}
    p_r = \dot{r}, \qquad 
    p_\varphi = r^{2}\dot{\varphi}, \qquad
    \dot{p}_r = \frac{r^{3}\dot{\varphi}^{2} - 1}{r^{2}},
\end{equation}
and rewrite $\dot{r}$, $r$, and $\dot{\varphi}$ in terms of their 
quasi-Keplerian parametrization \eqref{rqk}–\eqref{Kepler_quation}.  No canonical transformation is needed here, since harmonic and EOB coordinates start to differ only at 2PN order.

We observe that this procedure yields two distinct sources of contributions to the 
nonlocal spin–orbit Hamiltonian at 5.5PN:
\begin{itemize}
    \item[(i)] a contribution originating from the 1.5PN (linear-in-spin) terms of the 
    quasi-Keplerian parametrization, when these are applied to the nonlocal 4PN 
    component of the EOB potentials;
    \item[(ii)] a contribution arising directly from the nonlocal spin–orbit part of the 
    Hamiltonian itself, for which, being already of 5.5PN order, it is sufficient to 
    employ the Keplerian parametrization.
\end{itemize}

Expanding in eccentricity up to $\mathcal{O}(e_t^8)$ and performing the orbital 
average, we finally obtain the explicit expression of the 5.5PN nonlocal effective 
Delaunay Hamiltonian 
$\big\langle \delta H_{\text{eff,\,nonloc}}^{\text{LO}} \big\rangle$.  
The complete result is provided in electronic form in the supplementary file 
accompanying this paper~\cite{suppmat}.

\subsection{Nonlocal gyro-gravitomagnetic functions in full gauge generality at 5.5PN}
By PN-expanding the energy map \eqref{energy_map_rescaled} and explicitly 
separating its local and nonlocal contributions, it is straightforward to verify 
that, at leading nonlocal order, one simply has 
$\delta H_{\text{EOB,\,nonloc}}^{\text{LO}}
 = \delta H_{\text{eff,\,nonloc}}^{\text{LO}}$.
 
We can thus partially fix the coefficients of our ansatz for  $g_S^{5.5\text{PN}, \text{nonloc}}$ and $g_{S_*}^{5.5\text{PN}, \text{nonloc}}$ by 
directly identifying the two Delaunay-averaged Hamiltonians,
\begin{equation} \label{eq:matching_nl}
    \left\langle \delta H_\text{nonloc}^\text{LO} \right\rangle = \left\langle \delta H_\text{eff,\,nonloc}^\text{LO} \right\rangle.
\end{equation}
More precisely, the 5.5PN linear-in-spin sector of Eq.~\eqref{eq:matching_nl} 
yields 10 equations involving the 30 coefficients of 
$g_S^{5.5\text{PN},\text{nonloc}}$, and an additional 10 equations involving the 
corresponding 30 coefficients of $g_{S_*}^{5.5\text{PN},\text{nonloc}}$.  
Any solution of these two systems provides relations among the coefficients which, 
when substituted into the original ansatz, determine the gauge-unfixed expressions 
of the two gyro-gravitomagnetic functions. An explicit example of such a solution 
is provided in the supplementary material~\cite{suppmat}.

Our result for the gauge-general expressions for the nonlocal 5.5PN components of 
$g_S$ and $g_{S_*}$ reads
\begin{widetext}
\begin{subequations}
\label{eq:gfunctions_nl}
\begin{align}
\label{gs_nonlocal_5.5}
&g_S^{5.5\text{PN}, \text{nonloc}}=u^4 \left[\nu \left(-\frac{64}{5} - \frac{1168}{15}\gamma_E - \frac{464}{3}\log 2\right) - \frac{584}{15}\nu \log u \right] + \dot{p}_r^2 \left[ - g^{\text{N}^4\text{LO,nl}}_2 + \nu \left(\frac{25\,564}{15} - 208 \gamma_E\right.\right.\cr&\left.\left. + \frac{65\,488}{15}\log 2 - \frac{23\,328}{5}\log 3 \right) + \left(-104 \nu + g^{\text{N}^4\text{LO,nl}}_{2,\text{log}}\right)\log u \right]+ \frac{\dot{p}_r^4}{u^4}\left[ 8 g^{\text{N}^4\text{LO,nl}}_2 - g^{\text{N}^4\text{LO,nl}}_4 - \frac{1}{3} g^{\text{N}^4\text{LO,nl}}_5\right.\cr&\left. - \frac{10}{3} g^{\text{N}^4\text{LO,nl}}_{2,\text{log}} + \nu\left(-\frac{58\,102}{5} + 1664 \gamma_E - \frac{7\,926\,272}{45}\log 2 + 124\,902 \log 3 \right)\right. \left. + \left(832\nu - 8 g^{\text{N}^4\text{LO,nl}}_{2,\text{log}} + g^{\text{N}^4\text{LO,nl}}_{4,\text{log}}\right.\right.\cr&\left.\left. + \frac{1}{3} g^{\text{N}^4\text{LO,nl}}_{5,\text{log}}\right)\log u \right] + \frac{\dot{p}_r^6}{u^8}\left[-\frac{376}{5} g^{\text{N}^4\text{LO,nl}}_2 + \frac{47}{5} g^{\text{N}^4\text{LO,nl}}_4 + \frac{32}{15} g^{\text{N}^4\text{LO,nl}}_5 - g^{\text{N}^4\text{LO,nl}}_7 - \frac{1}{5} g^{\text{N}^4\text{LO,nl}}_8 - \frac{1}{5} g^{\text{N}^4\text{LO,nl}}_9 \right.\cr
&\left. + \frac{190}{3} g^{\text{N}^4\text{LO,nl}}_{2,\text{log}} - 4 g^{\text{N}^4\text{LO,nl}}_{4,\text{log}} - \frac{4}{5} g^{\text{N}^4\text{LO,nl}}_{5,\text{log}} + \nu \left(\frac{2\,724\,086}{25} - \frac{78\,208}{5}\gamma_E + \frac{80\,681\,472}{25}\log 2 - \frac{8\,387\,631}{5}\log 3\right.\right.\cr&\left.\left. - \frac{3\,015\,625}{9}\log 5 \right) + \left(-\frac{39\,104}{5}\nu + \frac{376}{5} g^{\text{N}^4\text{LO,nl}}_{2,\text{log}} - \frac{47}{5} g^{\text{N}^4\text{LO,nl}}_{4,\text{log}} - \frac{32}{15} g^{\text{N}^4\text{LO,nl}}_{5,\text{log}} + g^{\text{N}^4\text{LO,nl}}_{7,\text{log}} + \frac{1}{5} g^{\text{N}^4\text{LO,nl}}_{8,\text{log}}\right.\right.\cr&\left.\left. + \frac{1}{5} g^{\text{N}^4\text{LO,nl}}_{9,\text{log}}\right)\log u \right] + p_r^8 \left(g^{\text{N}^4\text{LO,nl}}_{11} - g^{\text{N}^4\text{LO,nl}}_{11,\text{log}}\log u\right)+ \frac{\dot{p}_r^8}{u^{12}} \left[712 g^{\text{N}^4\text{LO,nl}}_2 - 89 g^{\text{N}^4\text{LO,nl}}_4 - \frac{416}{21} g^{\text{N}^4\text{LO,nl}}_5\right.\cr&\left. + \frac{69}{7} g^{\text{N}^4\text{LO,nl}}_7 + \frac{61}{35} g^{\text{N}^4\text{LO,nl}}_8 + \frac{8}{7} g^{\text{N}^4\text{LO,nl}}_9 - g^{\text{N}^4\text{LO,nl}}_{11} - \frac{1}{7} g^{\text{N}^4\text{LO,nl}}_{12}  - \frac{3}{35} g^{\text{N}^4\text{LO,nl}}_{13} - \frac{1}{7} g^{\text{N}^4\text{LO,nl}}_{14} - \frac{18\,982}{21} g^{\text{N}^4\text{LO,nl}}_{2,\text{log}}\right.\cr&\left. + \frac{1594}{21} g^{\text{N}^4\text{LO,nl}}_{4,\text{log}} + \frac{4964}{315} g^{\text{N}^4\text{LO,nl}}_{5,\text{log}} - \frac{30}{7} g^{\text{N}^4\text{LO,nl}}_{7,\text{log}} - \frac{24}{35} g^{\text{N}^4\text{LO,nl}}_{8,\text{log}} - \frac{2}{5} g^{\text{N}^4\text{LO,nl}}_{9,\text{log}} + \nu\left(-\frac{524\,132\,537}{525} + 148\,096 \gamma_E \right.\right.\cr&\left.\left.- \frac{192\,215\,036\,288}{4725}\log 2 + \frac{22\,292\,460\,117}{1400}\log 3 + \frac{11\,346\,546\,875}{1512}\log 5 \right) + \left(74\,048 \nu - 712 g^{\text{N}^4\text{LO,nl}}_{2,\text{log}} + 89 g^{\text{N}^4\text{LO,nl}}_{4,\text{log}}\right.\right.\cr&\left.\left. + \frac{416}{21} g^{\text{N}^4\text{LO,nl}}_{5,\text{log}} - \frac{69}{7} g^{\text{N}^4\text{LO,nl}}_{7,\text{log}} - \frac{61}{35} g^{\text{N}^4\text{LO,nl}}_{8,\text{log}} - \frac{8}{7} g^{\text{N}^4\text{LO,nl}}_{9,\text{log}} + g^{\text{N}^4\text{LO,nl}}_{11,\text{log}} + \frac{1}{7} g^{\text{N}^4\text{LO,nl}}_{12,\text{log}} + \frac{3}{35} g^{\text{N}^4\text{LO,nl}}_{13,\text{log}} + \frac{1}{7} g^{\text{N}^4\text{LO,nl}}_{14,\text{log}} \right)\log u \right] \cr &+ p_r^6 \left[ u \left(g^{\text{N}^4\text{LO,nl}}_7 - g^{\text{N}^4\text{LO,nl}}_{7,\text{log}}\log u\right) + \frac{\dot{p}_r^2}{u^3}\left(g^{\text{N}^4\text{LO,nl}}_{12} - g^{\text{N}^4\text{LO,nl}}_{12,\text{log}}\log u\right) \right] + p_r^4 \left[ u^2 \left(g^{\text{N}^4\text{LO,nl}}_4 - g^{\text{N}^4\text{LO,nl}}_{4,\text{log}}\log u\right) \right.\cr&\left.+ \frac{\dot{p}_r^2}{u^2}\left(g^{\text{N}^4\text{LO,nl}}_8 - g^{\text{N}^4\text{LO,nl}}_{8,\text{log}}\log u\right) + \frac{\dot{p}_r^4}{u^6}\left(g^{\text{N}^4\text{LO,nl}}_{13} - g^{\text{N}^4\text{LO,nl}}_{13,\text{log}}\log u\right) \right] + p_r^2 \left[ u^3 \left(g^{\text{N}^4\text{LO,nl}}_2 - g^{\text{N}^4\text{LO,nl}}_{2,\text{log}}\log u\right) \right.\cr&\left.+ \frac{\dot{p}_r^2}{u}\left(g^{\text{N}^4\text{LO,nl}}_5 - g^{\text{N}^4\text{LO,nl}}_{5,\text{log}}\log u\right) + \frac{\dot{p}_r^4}{u^5}\left(g^{\text{N}^4\text{LO,nl}}_9 - g^{\text{N}^4\text{LO,nl}}_{9,\text{log}}\log u\right) + \frac{\dot{p}_r^6}{u^9}\left(g^{\text{N}^4\text{LO,nl}}_{14} - g^{\text{N}^4\text{LO,nl}}_{14,\text{log}}\log u\right) \right],\\
\cr 
&g_{S_*}^{5.5\text{PN}, \text{nonloc}}=  u^4 \left[\nu \left(-\frac{48}{5} - 48 \gamma_E - \frac{1\,456}{15}\log 2 \right) - 24 \nu \log u \right] + p_r^8 \left(g^{\text{N}^4\text{LO,nl}}_{*11} - g^{\text{N}^4\text{LO,nl}}_{*11,\text{log}} \log u \right) \cr
&+ \dot{p}_r^2 \left[- g^{\text{N}^4\text{LO,nl}}_{*2} 
+ \nu \left(\frac{17\,512}{15} - \frac{512}{5}\gamma_E + \frac{46\,976}{15}\log 2 - \frac{16\,038}{5}\log 3 \right) 
+ \left(-\frac{256}{5}\nu + g^{\text{N}^4\text{LO,nl}}_{*2,\text{log}} \right)\log u \right] \cr
&+ \frac{\dot{p}_r^4}{u^4} \left[ 8 g^{\text{N}^4\text{LO,nl}}_{*2} - \frac{10}{3} g^{\text{N}^4\text{LO,nl}}_{*2,\text{log}} - g^{\text{N}^4\text{LO,nl}}_{*4} - \frac{1}{3} g^{\text{N}^4\text{LO,nl}}_{*5}  + \nu \left(-\frac{123\,688}{15} + \frac{4\,096}{5}\gamma_E - \frac{373\,024}{3}\log 2 + 87\,480 \log 3 \right) \right.\cr
&\left.
+ \left(\frac{2\,048}{5}\nu - 8 g^{\text{N}^4\text{LO,nl}}_{*2,\text{log}} + g^{\text{N}^4\text{LO,nl}}_{*4,\text{log}} + \frac{1}{3} g^{\text{N}^4\text{LO,nl}}_{*5,\text{log}}\right)\log u \right] + \frac{\dot{p}_r^8}{u^{12}} \left[- g^{\text{N}^4\text{LO,nl}}_{*11} - \frac{1}{7} g^{\text{N}^4\text{LO,nl}}_{*12} - \frac{3}{35} g^{\text{N}^4\text{LO,nl}}_{*13} - \frac{1}{7} g^{\text{N}^4\text{LO,nl}}_{*14}\right.\cr
&\left. + 712 g^{\text{N}^4\text{LO,nl}}_{*2} - \frac{18\,982}{21} g^{\text{N}^4\text{LO,nl}}_{*2,\text{log}} \right.  - 89 g^{\text{N}^4\text{LO,nl}}_{*4} + \frac{1\,594}{21} g^{\text{N}^4\text{LO,nl}}_{*4,\text{log}} - \frac{416}{21} g^{\text{N}^4\text{LO,nl}}_{*5} + \frac{4\,964}{315} g^{\text{N}^4\text{LO,nl}}_{*5,\text{log}} + \frac{69}{7} g^{\text{N}^4\text{LO,nl}}_{*7}\cr
& - \frac{30}{7} g^{\text{N}^4\text{LO,nl}}_{*7,\text{log}}  + \frac{61}{35} g^{\text{N}^4\text{LO,nl}}_{*8} - \frac{24}{35} g^{\text{N}^4\text{LO,nl}}_{*8,\text{log}} + \frac{8}{7} g^{\text{N}^4\text{LO,nl}}_{*9} - \frac{2}{5} g^{\text{N}^4\text{LO,nl}}_{*9,\text{log}} + \nu \left(-\frac{5\,043\,336}{7} + \frac{364\,544}{5}\gamma_E\right.\cr
&\left. - \frac{135\,717\,590\,272}{4\,725}\log 2 + \frac{1\,576\,308\,681}{140}\log 3 + \frac{3\,964\,671\,875}{756}\log 5 \right) + \left(\frac{182\,272}{5}\nu + g^{\text{N}^4\text{LO,nl}}_{*11,\text{log}} + \frac{1}{7} g^{\text{N}^4\text{LO,nl}}_{*12,\text{log}} \right.\cr
&\left.+ \frac{3}{35} g^{\text{N}^4\text{LO,nl}}_{*13,\text{log}} + \frac{1}{7} g^{\text{N}^4\text{LO,nl}}_{*14,\text{log}} - 712 g^{\text{N}^4\text{LO,nl}}_{*2,\text{log}} \right. \left.\left. + 89 g^{\text{N}^4\text{LO,nl}}_{*4,\text{log}} + \frac{416}{21} g^{\text{N}^4\text{LO,nl}}_{*5,\text{log}} - \frac{69}{7} g^{\text{N}^4\text{LO,nl}}_{*7,\text{log}} - \frac{61}{35} g^{\text{N}^4\text{LO,nl}}_{*8,\text{log}} - \frac{8}{7} g^{\text{N}^4\text{LO,nl}}_{*9,\text{log}}\right)\log u \right] \cr
&+ \frac{\dot{p}_r^6}{u^8} \left[-\frac{376}{5} g^{\text{N}^4\text{LO,nl}}_{*2} + \frac{190}{3} g^{\text{N}^4\text{LO,nl}}_{*2,\text{log}} + \frac{47}{5} g^{\text{N}^4\text{LO,nl}}_{*4} - 4 g^{\text{N}^4\text{LO,nl}}_{*4,\text{log}} + \frac{32}{15} g^{\text{N}^4\text{LO,nl}}_{*5} - \frac{4}{5} g^{\text{N}^4\text{LO,nl}}_{*5,\text{log}} \right.  - g^{\text{N}^4\text{LO,nl}}_{*7} - \frac{1}{5} g^{\text{N}^4\text{LO,nl}}_{*8}\cr
&- \frac{1}{5} g^{\text{N}^4\text{LO,nl}}_{*9} + \nu \left(\frac{5\,839\,172}{75} - \frac{192\,512}{25}\gamma_E + \frac{1\,537\,664\,096}{675}\log 2 - \frac{5\,911\,461}{5}\log 3 - \frac{6\,296\,875}{27}\log 5 \right)\cr
&+ \left(-\frac{96\,256}{25}\nu + \frac{376}{5} g^{\text{N}^4\text{LO,nl}}_{*2,\text{log}} - \frac{47}{5} g^{\text{N}^4\text{LO,nl}}_{*4,\text{log}} - \frac{32}{15} g^{\text{N}^4\text{LO,nl}}_{*5,\text{log}} + g^{\text{N}^4\text{LO,nl}}_{*7,\text{log}} + \frac{1}{5} g^{\text{N}^4\text{LO,nl}}_{*8,\text{log}} + \frac{1}{5} g^{\text{N}^4\text{LO,nl}}_{*9,\text{log}} \right)\log u \Bigg] \cr
&+ p_r^6 \left[ \frac{\dot{p}_r^2}{u^3}\left(g^{\text{N}^4\text{LO,nl}}_{*12} - g^{\text{N}^4\text{LO,nl}}_{*12,\text{log}}\log u\right) + u\left(g^{\text{N}^4\text{LO,nl}}_{*7} - g^{\text{N}^4\text{LO,nl}}_{*7,\text{log}}\log u\right) \right]+ p_r^4 \left[ \frac{\dot{p}_r^4}{u^6}\left(g^{\text{N}^4\text{LO,nl}}_{*13} - g^{\text{N}^4\text{LO,nl}}_{*13,\text{log}}\log u\right)\right. \cr
&\left.+ u^2\left(g^{\text{N}^4\text{LO,nl}}_{*4} - g^{\text{N}^4\text{LO,nl}}_{*4,\text{log}}\log u\right) + \frac{\dot{p}_r^2}{u^2}\left(g^{\text{N}^4\text{LO,nl}}_{*8} - g^{\text{N}^4\text{LO,nl}}_{*8,\text{log}}\log u\right)\right]+ p_r^2 \left[ \frac{\dot{p}_r^6}{u^9}\left(g^{\text{N}^4\text{LO,nl}}_{*14} - g^{\text{N}^4\text{LO,nl}}_{*14,\text{log}}\log u\right) \right.\cr
&\left.+ u^3\left(g^{\text{N}^4\text{LO,nl}}_{*2} - g^{\text{N}^4\text{LO,nl}}_{*2,\text{log}}\log u\right) + \frac{\dot{p}_r^2}{u}\left(g^{\text{N}^4\text{LO,nl}}_{*5} - g^{\text{N}^4\text{LO,nl}}_{*5,\text{log}}\log u\right) + \frac{\dot{p}_r^4}{u^5}\left(g^{\text{N}^4\text{LO,nl}}_{*9} - g^{\text{N}^4\text{LO,nl}}_{*9,\text{log}}\log u\right)\right]. 
\label{gsstar_nonlocal_5.5}
\end{align}
\end{subequations}
\end{widetext}
The remaining coefficients in these expressions should be regarded as gauge coefficients, meaning that they do not affect physical observables, or more precisely, their determination at the 5.5PN order. By 
choosing their values appropriately, one can obtain the corresponding forms of 
$g_{S}^{5.5\text{PN},\text{nonloc}}$ and 
$g_{S_*}^{5.5\text{PN},\text{nonloc}}$ in any desired spin gauge.

\section{Local spin-orbit dynamics in full gauge generality at 5.5PN}
\label{sec:  loc}
We now turn to the local contribution to the 5.5PN conservative spin–orbit 
dynamics. Following Refs.~\cite{Antonelli:2020ybz,Khalil_2021,placidi}, the physical quantity we work with in this case is the conservative scattering angle.

\subsection{Conservative scattering angle: 5.5PN linear-in-spin terms}

The scattering angle of a two-body gravitational system is a gauge-invariant 
quantity defined, in the context of unbound motion, as the deflection experienced 
by one body as a result of its interaction with the other. It is now well known 
that the conservative, local-in-time part of the scattering angle can be 
analytically continued from unbound to bound dynamics by replacing the scattering 
states with the corresponding bound states of the two-body system 
\cite{SA1Damour_2016,SA2_Damour_2018}. The scattering angle can therefore be used as a 
powerful bridge between scattering results and bound-state observables, allowing 
one to fully exploit the fact that it can be computed directly through both 
perturbative and numerical methods.


Regarding the PN determination of the linear-in-spin contributions to the conservative scattering angle, Ref.~\cite{Khalil_2021}, which adapted the general strategy of Refs.~\cite{Bini:2019nra,Bini:2020wpo,Bini:2020nsb,Bini:2020hmy} to spinning systems, extended the 4.5PN-accurate linear-in-spin result of Ref.~\cite{Antonelli_2020} through the computation of the local 5.5PN terms [see Eq.~(3.31) therein]. The resulting expression is determined up to a single 
undetermined coefficient in the $\nu^{2}$ sector, $\mathsf{X}_{59}^{\nu^{2}}$, 
whose value awaits input from second-order GSF calculations.

In the absence of a fully determined result, our aim here is to use the 
available scattering-angle components to derive, in gauge-general form, the 
corresponding 5.5PN local contributions to $g_S$ and $g_{S_*}$.

To this end, we must compute the conservative scattering angle associated with our 
effective Hamiltonian. In general, the relation between the two is given by
\begin{equation}    \label{scat}
    \chi_{\text{eff}} \equiv -\pi - 2 \int_{0}^{u_{\text{max}}} \frac{du}{u^2} \, \frac{\partial}{\partial p_{\varphi}} p_r(\hat{E}_{\mathrm{eff}}, p_{\varphi}, u),
\end{equation}
where $p_{r}(\hat{E}_{\mathrm{eff}}, p_{\varphi}, u)$ is obtained from the 
iterative solution to the Hamilton–Jacobi equation in the effective problem, $u_{\max}$ denotes the largest root of 
$p_{r}(\hat{E}_{\mathrm{eff}}, p_{\varphi}, u)=0$, and 
$\hat{E}_{\mathrm{eff}}$ is the rescaled effective energy, related to the 
effective Hamiltonian through the energy-conservation condition 
$\hat{H}_{\mathrm{eff}}=\hat{E}_{\mathrm{eff}}$.  
The subscript “eff’’ on the scattering angle is included to emphasize that the 
quantity is computed from $\hat{H}_{\mathrm{eff}}$.

To compute the local contribution to the scattering angle, the integral 
\eqref{scat} must be evaluated using only the local part of the effective 
Hamiltonian, both in the EOB potentials and in the gyro-gravitomagnetic 
functions. As for the local 5.5PN contributions to the latter, and following the 
same strategy adopted in the previous section for the nonlocal terms, we 
introduce the most general ansatz compatible with this PN order, namely
\begin{widetext}
\begin{align}
   &g_S^{5.5\text{PN,loc}}=\frac{1}{c^8}\bigg( g^{\text{N}^4\text{LO}}_1 \; p^8 + 
g^{\text{N}^4\text{LO}}_2 \; p^6 p_r^2 + 
g^{\text{N}^4\text{LO}}_3 \; p^4 p_r^4 + 
g^{\text{N}^4\text{LO}}_4 \; p^2 \, p_r^6 + 
g^{\text{N}^4\text{LO}}_5 \; p_r^8 + 
g^{\text{N}^4\text{LO}}_6 \; p^6 u \cr &\hspace{0.5cm}
+g^{\text{N}^4\text{LO}}_7 \; p^4 p_r^2 u + 
g^{\text{N}^4\text{LO}}_8 \; p^2 \, p_r^4 u + 
g^{\text{N}^4\text{LO}}_9 \; p_r^6 u + 
g^{\text{N}^4\text{LO}}_{10} \; p^4 u^2 + 
g^{\text{N}^4\text{LO}}_{11} \; p^2 \, p_r^2 u^2 + 
g^{\text{N}^4\text{LO}}_{12} \; p_r^4 u^2\cr &\hspace{0.5cm}
+ g^{\text{N}^4\text{LO}}_{13} \; p^2 u^3 + 
g^{\text{N}^4\text{LO}}_{14} \; p_r^2 u^3 + 
g^{\text{N}^4\text{LO}}_{15} \; u^4\bigg),
\label{local_ansatz}
\end{align}    
\end{widetext}
where we have introduced a set of 15 coefficients, $g^{\text{N}^4\text{LO}}_{n}$.  
An analogous ansatz, involving another set of 15 coefficients 
$g^{\text{N}^4\text{LO}}_{*n}$, is used for $g_{S_*}^{5.5\text{PN,loc}}$.  
The different structure of this ansatz, as compared to Eq.~\eqref{ansatz_gs}, 
reflects the fact that in the local sector we neither perform a small-eccentricity 
expansion nor need to account for logarithmic contributions.

At this stage, the 5.5PN determination of $\chi_{\rm eff}$ is obtained by 
following the same procedure outlined in Sec.~IIIB of Ref.~\cite{placidi}, 
extending the simultaneous expansion of Eq.~\eqref{scat} in powers of 
$1/p_\varphi$ and $1/c$ up to orders $1/p_\varphi^{5}$ and $1/c^{10}$, 
respectively.

The resulting 5.5PN spin–orbit contribution to the scattering angle reads
\begin{align}
    \chi_{\text{eff}}^{\text{5.5PN,\,SO}}
    =&\;
    S\, \chi_{S}^{\text{5.5PN}}
        (\hat{E}_{\text{eff}}, p_{\varphi}, g^{\text{N}^4\text{LO}}_{n})
\cr &
    + S_{*}\, \chi_{S_*}^{\text{5.5PN}}
        (\hat{E}_{\text{eff}}, p_{\varphi}, g^{\text{N}^4\text{LO}}_{*n}) ,
\end{align}
which exhibits two distinct linear-in-spin contributions, each depending on the 
coefficients of the general ansatz \eqref{local_ansatz} and its 
$g_{S_*}^{\text{5.5PN,\,loc}}$ analogue.  
The explicit expressions for both components are provided in the supplementary 
material~\cite{suppmat}.

\subsection{Local gyro-gravitomagnetic functions in full gauge generality at 5.5PN}

We can now proceed to constrain the coefficients of our general ansatz by imposing 
the matching condition
\begin{equation} \label{eq:sa_matching}
    \chi_{\text{eff}}^{\text{5.5PN,\,SO}}
    = \chi_{\text{Khalil}}^{\text{5.5PN,\,SO}},
\end{equation}
which follows from the gauge invariance of the scattering angle.  
Here $\chi_{\text{Khalil}}^{\text{5.5PN,\,SO}}$ denotes the local linear-in-spin 
5.5PN contribution to the scattering angle, given in Eq.~(3.31) of 
Ref.~\cite{Khalil_2021}.


From the $S$ and $S_{*}$ components of Eq.~\eqref{eq:sa_matching}, we obtain two 
sets of five equations relating the coefficients of the local ansatz. An explicit 
example of a solution to these systems is provided in the supplementary material 
\cite{suppmat}.

Substituting such a solution into our general ansatz for 
$g_{S}^{\text{5.5PN,\,loc}}$ and $g_{S_*}^{\text{5.5PN,\,loc}}$, we finally obtain
\begin{widetext}
\begin{subequations}
\label{eq:gfunctions_loc}
\begin{align}    &g_S^{5.5\text{PN,loc}}=g_1^{\text{N}^4\text{LO}}\,p^8 + \bigg(-\frac{27 \nu}{64} + \frac{99 \nu^2}{128} + \frac{945 \nu^3}{256} + \frac{189 \nu^4}{32} - 9\,g_1^{\text{N}^4\text{LO}} - 3\,g_2^{\text{N}^4\text{LO}} - \frac{9}{5}g_3^{\text{N}^4\text{LO}} - \frac{9}{7}g_4^{\text{N}^4\text{LO}} \bigg)p_r^8\cr
    &+ \bigg(g_2^{\text{N}^4\text{LO}}\,p_r^2 + g_6^{\text{N}^4\text{LO}}\,u \bigg)p^6 + \bigg( \frac{16}{3}g_2^{\text{N}^3\text{LO}} + 6\,g_4^{\text{N}^3\text{LO}} + \frac{41}{7}g_7^{\text{N}^3\text{LO}} - \frac{467 \nu}{64} - \frac{1\,077 \nu^2}{128} - \frac{3\,607 \nu^3}{256} + \frac{1\,577 \nu^4}{160}\cr
    & - 65\,g_1^{\text{N}^4\text{LO}} - 11\,g_2^{\text{N}^4\text{LO}} - \frac{17}{5}g_3^{\text{N}^4\text{LO}} - g_4^{\text{N}^4\text{LO}} - \frac{64}{5}g_6^{\text{N}^4\text{LO}} - \frac{16}{5}g_7^{\text{N}^4\text{LO}} - \frac{8}{5}g_8^{\text{N}^4\text{LO}} \bigg)p_r^6 u + \bigg(g_3^{\text{N}^4\text{LO}}\,p_r^4\cr
    & + g_7^{\text{N}^4\text{LO}}\,p_r^2 u + g_{10}^{\text{N}^4\text{LO}}\,u^2 \bigg)p^4 + \bigg[ 6\,g_2^{\text{N}^2\text{LO}} + \frac{64}{3}g_2^{\text{N}^3\text{LO}} + \frac{79}{20}g_4^{\text{N}^2\text{LO}} + \frac{88}{5}g_4^{\text{N}^3\text{LO}} + \frac{3}{2}g_5^{\text{N}^3\text{LO}} + \frac{377}{28}g_7^{\text{N}^3\text{LO}}\cr
    &+ \frac{7}{4}g_8^{\text{N}^3\text{LO}} + \bigg( \frac{205}{4} + 6\,g_2^{\text{N}^2\text{LO}} + 8\,g_2^{\text{N}\text{LO}} + \frac{12}{5}g_4^{\text{N}^2\text{LO}} \bigg)\nu + \bigg( \frac{30\,541}{128} - 6\,g_2^{\text{N}\text{LO}} \bigg)\nu^2 - \frac{60\,813 \nu^3}{256} - \frac{107 \nu^4}{32}\cr
    &- 95\,g_1^{\text{N}^4\text{LO}} - 8\,g_2^{\text{N}^4\text{LO}} - g_3^{\text{N}^4\text{LO}} - 41\,g_6^{\text{N}^4\text{LO}} - 5\,g_7^{\text{N}^4\text{LO}} - g_8^{\text{N}^4\text{LO}} - \frac{35}{3}g_{10}^{\text{N}^4\text{LO}} - \frac{7}{3}g_{11}^{\text{N}^4\text{LO}} \bigg]p_r^4 u^2 + \bigg[ \frac{91}{6}g_2^{\text{N}^2\text{LO}}\cr
    &+ \frac{44}{5}g_2^{\text{N}^3\text{LO}} + \frac{461}{60}g_2^{\text{N}\text{LO}} + \frac{323}{50}g_4^{\text{N}^2\text{LO}} + \frac{9}{5}g_4^{\text{N}^3\text{LO}} + \frac{53}{20}g_5^{\text{N}^2\text{LO}} + \frac{59}{10}g_5^{\text{N}^3\text{LO}} - \frac{51}{28}g_7^{\text{N}^3\text{LO}} + \frac{87}{20}g_8^{\text{N}^3\text{LO}}\cr
    &+ \frac{1}{5}g_9^{\text{N}^3\text{LO}} + \bigg( \frac{1\,869\,859}{1\,600} - 4\,g_2^{\text{N}^2\text{LO}} + \frac{28}{5}g_2^{\text{N}\text{LO}} - 6\,g_4^{\text{N}^2\text{LO}} + 3\,g_5^{\text{N}^2\text{LO}} - \frac{22\,301 \pi^2}{256} \bigg)\nu + \bigg( \frac{877\,631}{960} \cr
    &- g_2^{\text{N}\text{LO}} - \frac{1\,087 \pi^2}{64} \bigg)\nu^2 - \frac{12\,579 \nu^3}{128} + \frac{3 \nu^4}{4} - 30\,g_1^{\text{N}^4\text{LO}} - g_2^{\text{N}^4\text{LO}} - 21\,g_6^{\text{N}^4\text{LO}} - g_7^{\text{N}^4\text{LO}} - 13\,g_{10}^{\text{N}^4\text{LO}} - g_{11}^{\text{N}^4\text{LO}}\cr
    &- 6\,g_{13}^{\text{N}^4\text{LO}} \bigg]p_r^2 u^3 + \bigg[ \frac{1}{2}g_2^{\text{N}^2\text{LO}} + \frac{2}{5}g_2^{\text{N}^3\text{LO}} + \frac{101}{20}g_2^{\text{N}\text{LO}} - \frac{369}{200}g_4^{\text{N}^2\text{LO}} + \frac{39}{20}g_5^{\text{N}^2\text{LO}} + \frac{9}{20}g_5^{\text{N}^3\text{LO}} + \frac{3}{56}g_7^{\text{N}^3\text{LO}}\cr
    &- \frac{3}{40}g_8^{\text{N}^3\text{LO}} + \frac{3}{5}g_9^{\text{N}^3\text{LO}} + \bigg( -\frac{4\,546\,811}{14\,400} - \frac{1}{2}g_2^{\text{N}^2\text{LO}} - \frac{4\,681}{180}g_2^{\text{N}\text{LO}} + \frac{3}{20}g_4^{\text{N}^2\text{LO}} - \frac{3}{4}g_5^{\text{N}^2\text{LO}} + \frac{62\,041 \pi^2}{1\,536}\cr
    &+ \frac{41 \pi^2 g_2^{\text{N}\text{LO}}}{48} \bigg)\nu + \bigg( \frac{313\,823}{5\,760} - \frac{1\,225 \pi^2}{384} - \frac{3 \mathsf{X}^{\nu^2}_{59}}{16} \bigg)\nu^2 - \frac{239 \nu^3}{256} - \frac{\nu^4}{32} - g_1^{\text{N}^4\text{LO}} - g_6^{\text{N}^4\text{LO}} - g_{10}^{\text{N}^4\text{LO}} \cr
    &- g_{13}^{\text{N}^4\text{LO}} \bigg]u^4 + \bigg( g_4^{\text{N}^4\text{LO}}\,p_r^6 + g_8^{\text{N}^4\text{LO}}\,p_r^4 u + g_{11}^{\text{N}^4\text{LO}}\,p_r^2 u^2 + g_{13}^{\text{N}^4\text{LO}}\,u^3 \bigg)p^2,\label{gs_local_gaugegeneral} \\
\cr
&g_{S_*}^{5.5\text{PN,loc}}=g_{*1}^{\text{N}^4\text{LO}}\,p^8 + \bigg( \frac{693}{256} + \frac{189}{64}\nu + \frac{1\,053}{256}\nu^2 + \frac{315}{64}\nu^3 + \frac{945}{256}\nu^4 - 9\,g_{*1}^{\text{N}^4\text{LO}} - 3\,g_{*2}^{\text{N}^4\text{LO}}- \frac{9}{5}g_{*3}^{\text{N}^4\text{LO}}\cr
& - \frac{9}{7}g_{*4}^{\text{N}^4\text{LO}} \bigg)\,p_r^8+ \bigg( g_{*2}^{\text{N}^4\text{LO}}\,p_r^2 + g_{*6}^{\text{N}^4\text{LO}}\,u \bigg)\,p^6 + \bigg( \frac{7\,245}{256} + \frac{16}{3}g_{*2}^{\text{N}3\text{LO}} + 6\,g_{*4}^{\text{N}3\text{LO}} + \frac{41}{7}g_{*7}^{\text{N}3\text{LO}} + \frac{1\,333}{64}\nu\cr
& + \frac{4\,533}{256}\nu^2 + \frac{3\,759}{320}\nu^3 + \frac{16\,077}{1\,280}\nu^4 - 65\,g_{*1}^{\text{N}^4\text{LO}} - 11\,g_{*2}^{\text{N}^4\text{LO}} - \frac{17}{5}g_{*3}^{\text{N}^4\text{LO}} - g_{*4}^{\text{N}^4\text{LO}} - \frac{64}{5}g_{*6}^{\text{N}^4\text{LO}} - \frac{16}{5}g_{*7}^{\text{N}^4\text{LO}} \cr
&- \frac{8}{5}g_{*8}^{\text{N}^4\text{LO}} \bigg)\,p_r^6\,u + \bigg( g_{*3}^{\text{N}^4\text{LO}}\,p_r^4 + g_{*7}^{\text{N}^4\text{LO}}\,p_r^2\,u + g_{*10}^{\text{N}^4\text{LO}}\,u^2 \bigg)\,p^4 + \bigg[ \frac{12\,663}{256} + 6\,g_{*2}^{\text{N}2\text{LO}} + \frac{64}{3}g_{*2}^{\text{N}3\text{LO}} + \frac{79}{20}g_{*4}^{\text{N}2\text{LO}}\cr
& + \frac{88}{5}g_{*4}^{\text{N}3\text{LO}} + \frac{3}{2}g_{*5}^{\text{N}3\text{LO}} + \frac{377}{28}g_{*7}^{\text{N}3\text{LO}} + \frac{7}{4}g_{*8}^{\text{N}3\text{LO}} + \bigg( \frac{3\,297}{64} + 6\,g_{*2}^{\text{N}2\text{LO}} + 8\,g_{*2}^{\text{N}\text{LO}} + \frac{12}{5}g_{*4}^{\text{N}2\text{LO}} \bigg)\nu + \bigg( \frac{19\,467}{256} \cr
&- 6\,g_{*2}^{\text{N}\text{LO}} \bigg)\nu^2 - \frac{11\,523}{64}\nu^3 - \frac{1\,377}{256}\nu^4 - 95\,g_{*1}^{\text{N}^4\text{LO}} - 8\,g_{*2}^{\text{N}^4\text{LO}} - g_{*3}^{\text{N}^4\text{LO}} - 41\,g_{*6}^{\text{N}^4\text{LO}} - 5\,g_{*7}^{\text{N}^4\text{LO}} - g_{*8}^{\text{N}^4\text{LO}}\cr
& - \frac{35}{3}g_{*10}^{\text{N}^4\text{LO}} - \frac{7}{3}g_{*11}^{\text{N}^4\text{LO}} \bigg]\,p_r^4\,u^2 + \bigg[ \frac{13\,577}{640} + \frac{91}{6}g_{*2}^{\text{N}2\text{LO}} + \frac{44}{5}g_{*2}^{\text{N}3\text{LO}} + \frac{461}{60}g_{*2}^{\text{N}\text{LO}} + \frac{323}{50}g_{*4}^{\text{N}2\text{LO}} + \frac{9}{5}g_{*4}^{\text{N}3\text{LO}} \cr
&+ \frac{53}{20}g_{*5}^{\text{N}2\text{LO}} + \frac{59}{10}g_{*5}^{\text{N}3\text{LO}} - \frac{51}{28}g_{*7}^{\text{N}3\text{LO}} + \frac{87}{20}g_{*8}^{\text{N}3\text{LO}} + \frac{1}{5}g_{*9}^{\text{N}3\text{LO}} + \bigg( \frac{27\,739}{80} - 4\,g_{*2}^{\text{N}2\text{LO}} + \frac{28}{5}g_{*2}^{\text{N}\text{LO}} - 6\,g_{*4}^{\text{N}2\text{LO}}\cr
& + 3\,g_{*5}^{\text{N}2\text{LO}} - \frac{4\,829\,\pi^2}{256} \bigg)\nu + \bigg( \frac{360\,199}{640} - g_{*2}^{\text{N}\text{LO}} - \frac{123\,\pi^2}{8} \bigg)\nu^2 - \frac{2\,931}{32}\nu^3 + \frac{171}{128}\nu^4 - 30\,g_{*1}^{\text{N}^4\text{LO}} - g_{*2}^{\text{N}^4\text{LO}}\cr
& - 21\,g_{*6}^{\text{N}^4\text{LO}} - g_{*7}^{\text{N}^4\text{LO}} - 13\,g_{*10}^{\text{N}^4\text{LO}} - g_{*11}^{\text{N}^4\text{LO}} - 6\,g_{*13}^{\text{N}^4\text{LO}} \bigg]\,p_r^2\,u^3 + \bigg[ -\frac{2\,103}{1\,280} + \frac{1}{2}g_{*2}^{\text{N}2\text{LO}} + \frac{2}{5}g_{*2}^{\text{N}3\text{LO}} + \frac{101}{20}g_{*2}^{\text{N}\text{LO}}\cr
& - \frac{369}{200}g_{*4}^{\text{N}2\text{LO}} + \frac{39}{20}g_{*5}^{\text{N}2\text{LO}} + \frac{9}{20}g_{*5}^{\text{N}3\text{LO}} + \frac{3}{56}g_{*7}^{\text{N}3\text{LO}} - \frac{3}{40}g_{*8}^{\text{N}3\text{LO}} + \frac{3}{5}g_{*9}^{\text{N}3\text{LO}} + \bigg( -\frac{9\,785}{192} - \frac{1}{2}g_{*2}^{\text{N}2\text{LO}} \cr
&- \frac{4\,681}{180}g_{*2}^{\text{N}\text{LO}} + \frac{3}{20}g_{*4}^{\text{N}2\text{LO}} - \frac{3}{4}g_{*5}^{\text{N}2\text{LO}} + \frac{26\,943\,\pi^2}{2\,048} + \frac{41}{48}g_{*2}^{\text{N}\text{LO}}\,\pi^2 \bigg)\nu + \bigg( \frac{9\,729}{1\,280} - \frac{123\,\pi^2}{64} - \frac{3}{16}\mathsf{X}^{\nu^2}_{59} \bigg)\nu^2\cr
& - \frac{57}{64}\nu^3 - \frac{15}{256}\nu^4 - g_{*1}^{\text{N}^4\text{LO}} - g_{*6}^{\text{N}^4\text{LO}} - g_{*10}^{\text{N}^4\text{LO}} - g_{*13}^{\text{N}^4\text{LO}} \bigg]\,u^4 + \bigg( g_{*4}^{\text{N}^4\text{LO}}\,p_r^6 + g_{*8}^{\text{N}^4\text{LO}}\,p_r^4\,u \cr
&+ g_{*11}^{\text{N}^4\text{LO}}\,p_r^2\,u^2 + g_{*13}^{\text{N}^4\text{LO}}\,u^3 \bigg)\,p^2 .
\label{gsstar_local_gaugegeneral}
\end{align}
\end{subequations}
\end{widetext}
The remaining ten coefficients in each of these expressions encode the freedom 
associated with the spin–gauge choice, in direct analogy with the coefficients 
appearing in Eqs.~\eqref{eq:gfunctions_nl}.

In accordance with the local/nonlocal decomposition of 
Eqs.~\eqref{eq:gyrogfunctions_5.5PN}, the full gauge-general expressions for 
$g_S^{\text{5.5PN}}$ and $g_{S_*}^{\text{5.5PN}}$ are obtained by combining 
Eqs.~\eqref{eq:gfunctions_nl} and \eqref{eq:gfunctions_loc}.  
Since these results constitute one of the main outcomes of this work, we provide 
their complete expressions in electronic form in the supplementary material 
\cite{suppmat}.

\section{Analytical consistency checks: Gauge invariant quantities}\label{sec: gauge-i}
In this section, we test the validity of our gauge-general results for the 
5.5PN gyro-gravitomagnetic functions. To this end, we compute the 5.5PN 
linear-in-spin contributions to two gauge-invariant quantities: the binding 
energy and the fractional periastron advance for quasi-circular orbits.  
A necessary consistency requirement for the gauge-general expressions of 
$g_S^{\text{5.5PN}}$ and $g_{S_*}^{\text{5.5PN}}$ is that any gauge-invariant 
observable must remain independent of the 30 gauge coefficients (10 local and 
20 nonlocal) entering these functions.


The rescaled effective binding energy has been computed up to 4.5PN order in 
Refs.~\cite{Antonelli_2020,placidi} and up to 5.5PN order in 
Ref.~\cite{Khalil_2021}. It is defined as
\begin{equation}
    E_b \equiv \hat{H}^{\text{circ}}_{\text{EOB}}(x) - \frac{1}{\nu},
    \label{definition: binding energy}
\end{equation}
where the rescaled EOB Hamiltonian \eqref{energy_map_rescaled}, evaluated along 
circular orbits, is expressed in terms of the frequency parameter 
$x \equiv \Omega_\varphi^{2/3}$, with $\Omega_\varphi$ the orbital frequency.  
Further details on the computation of $\hat{H}^{\text{circ}}_{\text{EOB}}(x)$ 
from the general-orbit EOB Hamiltonian can be found in Sec.~IV of 
Ref.~\cite{placidi}.

Turning to the fractional periastron advance per radial period, its definition is
\begin{equation}
    K = 1 + \frac{\Delta \Phi}{2\pi},
\end{equation}
and, more explicitly, for quasi-circular orbits one has
\begin{equation}
    K \equiv \left. \frac{\Omega_\varphi}{\Omega_r} \right|_{\text{circ}} 
= \left( \frac{\partial^2 \hat{H}_{\text{eff}}}{\partial r^2} \, \frac{\partial^2 \hat{H}_{\text{eff}}}{\partial p_r^2} \right)^{-1} 
\left. \frac{\partial \hat{H}_{\text{eff}}}{\partial p_\varphi} \right|_{\text{circ}},
\end{equation}
where the circular limit is taken only after the derivatives have been evaluated 
by subsequently rewriting all quantities in terms of the frequency parameter $x$.  
$K$ has been computed to 3.5PN order in Ref.~\cite{periaHinderer_2013} and to 
4.5PN order in Ref.~\cite{placidi}.

Computing the linear-in-spin 5.5PN contributions to $E_b$ and $K$ using our 
gauge-general expressions for $g_S$ and $g_{S_*}$ in the Hamiltonian, we find
\begin{widetext}
\begin{align}
\label{eq:Eb}
E_b^{5.5\rm PN,SO}=
& \,x^{13/2}  \bigg\{ S \bigg[-\frac{4\,725}{32} + \bigg( -\frac{1\,975\,415}{5\,184} + \frac{2\,425}{864} \pi^2 + \frac{5}{8} \mathsf{X}^{\nu^2}_{59} \bigg) \nu^2 + \frac{310\,795}{5\,184} \nu^3 + \frac{35}{1\,458} \nu^4 \cr
& + \nu \bigg( \frac{1\,411\,663}{640} + \frac{352}{3} \gamma_E - \frac{10\,325}{64} \pi^2 + \frac{2\,080}{9} \log 2 + \frac{176}{3} \log x  \bigg) \bigg]\cr
&+S_* \bigg[ -\frac{2\,835}{128} + \bigg( -\frac{275\,245}{3\,456} - \frac{205}{576} \pi^2 + \frac{5}{8} \mathsf{X}^{\nu^2}_{59} \bigg) \nu^2 + \frac{46\,765}{864} \nu^3 + \frac{875}{31\,104} \nu^4 \cr
&+ \nu \bigg( \frac{126\,715}{144} + \frac{160}{3} \gamma_E - \frac{102\,355}{1\,536} \pi^2 + \frac{992}{9} \log 2 + \frac{80}{3} \log x  \bigg) \bigg]   \bigg\},\\
\cr
K^{5.5\rm PN,SO}=& \,x^{11/2} \bigg\{ S \bigg[ 
-\frac{23\,625}{2} 
+ \bigg( -\frac{4\,266\,299}{288} + \frac{47\,263}{192} \pi^2 + \frac{15}{4} \mathsf{X}^{\nu^2}_{59} \bigg) \nu^2 
+ \frac{251\,375}{324} \nu^3 
- \frac{32}{81} \nu^4 \cr
& + \nu \bigg( \frac{3\,259\,681\,769}{86\,400} + \frac{112\,528}{45} \gamma_E 
- \frac{1\,345\,253}{1\,152} \pi^2 + \frac{48\,784}{45} \log 2 
+ \frac{25\,272}{5} \log 3 \cr
&- 1\,280 \log 2 + \frac{56\,264}{45} \log x  \bigg) 
\bigg] + S_* \bigg[ 
-\frac{42\,525}{8} 
+ \bigg( -\frac{326\,597}{32} + \frac{11\,931}{64} \pi^2 + \frac{15}{4} \mathsf{X}^{\nu^2}_{59} \bigg) \nu^2 \cr
&+ \frac{2\,737}{4} \nu^3 
- \frac{11}{27} \nu^4 + \nu \bigg( \frac{114\,877\,897}{5\,760} + \frac{24\,208}{15} \gamma_E 
- \frac{1\,679\,935}{3\,072} \pi^2 + \frac{8\,656}{15} \log 2 
+ \frac{17\,496}{5} \log 3\cr
& - 960 \log 2 + \frac{12\,104}{15} \log x  \bigg) 
\bigg] 
\bigg\}.
\end{align}
\end{widetext}
As expected, neither of the two expressions exhibits any residual dependence on 
the gauge coefficients. The only remaining parameter, 
$\mathsf{X}_{59}^{\nu^2}$, reflects the current incompleteness in the knowledge 
of the local linear-in-spin contribution to the 5.5PN scattering angle, as 
discussed in Sec.~\ref{sec: loc}.

We further observe that the binding energy \eqref{eq:Eb} is in perfect agreement 
with the 5.5PN result of Ref.~\cite{Khalil_2021}.

\section{Gauge flexibility of the gyro-gravitomagnetic functions}
\label{sec:gauge_fixing}

In this section, building on our gauge-general results for 
$g_{S}^{\text{5.5PN}}$ and $g_{S_*}^{\text{5.5PN}}$, we investigate how different 
spin gauge choices at this order compare with numerical data.  
In particular, following Ref.~\cite{placidi}, we focus on the DJS and 
$\overline{\rm DJS}$ spin gauges and examine their performance in reproducing 
the spin–orbit contribution to the circular-orbit binding energy.

\subsection{The DJS spin gauge}

The DJS spin gauge was first introduced in Ref.~\cite{NLO_effecrive_Damour_2008} 
and subsequently adopted in a wide range of works 
\cite{NagarDamour_2014, N2LOSO, placidi, N3LOAntonelli_2020,Antonelli_2020}.  
Its defining condition is straightforward: the gauge coefficients entering 
$g_S$ and $g_{S_*}$ must be fixed so as to eliminate any dependence on the total 
momentum $p^{2}$, or equivalently on the angular momentum $p_\varphi$.  
Imposing this gauge-fixing condition on our combined expressions 
\eqref{gs_nonlocal_5.5}+\eqref{gs_local_gaugegeneral} and \eqref{gsstar_nonlocal_5.5}+\eqref{gsstar_local_gaugegeneral} yields a system 
of equations that can be uniquely solved for all gauge coefficients.  
The corresponding solution, provided in the supplementary material 
\cite{suppmat}, gives the following 5.5PN expressions of the DJS 
gyro-gravitomagnetic functions:

\begin{widetext}
\begin{subequations}    
\begin{align}
&\left(g^{5.5\text{PN}}_S\right)_{\rm DJS}=\bigg[ \frac{31\,913}{128} \nu^2 - \frac{73\,547}{256} \nu^3 - \frac{107}{32} \nu^4 + \nu \bigg( \frac{3\,325\,823}{1920} - \frac{1\,270\,912 \log 2}{9} + \frac{437\,886 \log 3}{5} \bigg) \bigg] p_r^4 u^2 \cr
&\hspace{0.5cm}+ \bigg[ -\frac{2\,553}{128} \nu^2 - \frac{11\,397}{256} \nu^3 + \frac{1\,577}{160} \nu^4 + \nu \bigg( \frac{9\,768\,651}{3200} + \frac{353\,598\,464 \log 2}{225} - \frac{2\,517\,237 \log 3}{5}\cr
&\hspace{0.5cm} - \frac{3\,015\,625 \log 5}{9} \bigg) \bigg] p_r^6 u + \bigg[ \frac{99}{128} \nu^2 + \frac{945}{256} \nu^3 + \frac{189}{32} \nu^4 + \nu \bigg( \frac{32\,193\,611}{11200} - \frac{1\,797\,965\,696 \log 2}{189} \cr
&\hspace{0.5cm}- \frac{31\,129\,029 \log 3}{200} + \frac{6\,352\,671\,875 \log 5}{1512} \bigg) \bigg] p_r^8 + \bigg[ \bigg( \frac{198\,133}{192} - \frac{1087 \pi^2}{64} \bigg) \nu^2 - \frac{8\,259}{64} \nu^3 + \frac{3}{4} \nu^4 \cr
&\hspace{0.5cm}+ \nu \bigg( \frac{27\,198\,169}{9600} - 208 \gamma_E - \frac{22\,301 \pi^2}{256} + \frac{65\,488 \log 2}{15} - \frac{23\,328 \log 3}{5} \bigg) - 104 \nu \log u \bigg] p_r^2 u^3\cr
&\hspace{0.5cm} + \bigg[ \bigg( \frac{235\,111}{1152} - \frac{583 \pi^2}{96} - \frac{3}{16} \mathsf{X}^{\nu^2}_{59} \bigg) \nu^2 - \frac{413}{256} \nu^3 - \frac{1}{32} \nu^4 + \nu \bigg( -\frac{12\,015\,517}{28\,800} - \frac{1\,168}{15} \gamma_E \cr
&\hspace{0.5cm}+ \frac{62\,041 \pi^2}{1536} - \frac{464 \log 2}{3} \bigg) - \frac{584}{15} \nu \log u \bigg] u^4,\\
\cr
&\left(g^{5.5\text{PN}}_{S_*}\right)_{\rm DJS}=\bigg[ \frac{2\,525}{256} + \frac{12\,135}{256} \nu^2 - \frac{13\,905}{64} \nu^3 - \frac{1\,377}{256} \nu^4 + \nu \bigg( \frac{37\,463}{40} - \frac{1\,489\,312 \log 2}{15} + \frac{309\,096 \log 3}{5} \bigg) \bigg] p_r^4 u^2 \cr
&\hspace{0.5cm}+ \bigg[ \frac{3\,555}{256} - \frac{879}{256} \nu^2 - \frac{2\,391}{320} \nu^3 + \frac{16\,077}{1\,280} \nu^4 + \nu \bigg( \frac{795\,077}{400} + \frac{748\,718\,336 \log 2}{675} - \frac{1\,799\,901 \log 3}{5} \cr
&\hspace{0.5cm}- \frac{6\,296\,875 \log 5}{27} \bigg) \bigg] p_r^6 u+ \bigg[ \frac{693}{256} + \frac{1\,053}{256} \nu^2 + \frac{315}{64} \nu^3 + \frac{945}{256} \nu^4 + \nu \bigg( \frac{20\,870\,707}{11\,200} - \frac{4\,538\,197\,984 \log 2}{675} \cr
&\hspace{0.5cm}- \frac{2\,092\,959 \log 3}{28} + \frac{2\,226\,734\,375 \log 5}{756} \bigg) \bigg] p_r^8 + \bigg[ -\frac{27}{32} + \bigg( \frac{77\,201}{128} - \frac{123 \pi^2}{8} \bigg) \nu^2 - \frac{489}{4} \nu^3 + \frac{171}{128} \nu^4 \cr
&\hspace{0.5cm}+ \nu \bigg( \frac{1\,432\,861}{960} - \frac{512}{5} \gamma_E - \frac{4\,829 \pi^2}{256} + \frac{46\,976 \log 2}{15} - \frac{16\,038 \log 3}{5} \bigg) - \frac{256}{5} \nu \log u \bigg] p_r^2 u^3 \cr
&\hspace{0.5cm}+ \bigg[ -\frac{1\,701}{256} + \bigg( \frac{29\,081}{256} - \frac{123 \pi^2}{32} - \frac{3}{16} \mathsf{X}^{\nu^2}_{59} \bigg) \nu^2 - \frac{111}{64} \nu^3 - \frac{15}{256} \nu^4 + \nu \bigg( -\frac{1\,017}{20} - 48 \gamma_E\cr
&\hspace{0.5cm} + \frac{23\,663 \pi^2}{2\,048} - \frac{1\,456 \log 2}{15} \bigg) - 24 \nu \log u \bigg] u^4.
\end{align}
\end{subequations}
\end{widetext}
We observe that these expressions are in exact agreement with 
Eqs.~(2.24), (3.32), and (3.33) of Ref.~\cite{Khalil_2021}, where the computation 
was carried out entirely within the DJS spin gauge, i.e.\ by imposing this gauge 
choice directly at the level of the ansatz for the two gyro-gravitomagnetic 
functions.

\subsection{The $\overline{\rm DJS}$ spin gauge}
We now turn to a notable alternative to the DJS spin gauge, introduced in 
Ref.~\cite{antidjsRettegno_2020} and further explored in Ref.~\cite{placidi}, 
namely the $\overline{\text{DJS}}$ spin gauge.

The defining feature of this spin gauge is the requirement that, at each available 
PN order, the condition
\begin{equation}
    g_{S_*}^{\overline{\rm DJS}} \xrightarrow[\nu \to 0]{} g_{S_*}^{\rm K},
    \label{limit}
\end{equation}
be satisfied; that is, in the test-mass limit, $g_{S_*}$ must reduce to the 
gyro-gravitomagnetic function of a spinning test particle in a Kerr background, 
denoted by $g_{S_*}^{\rm K}$.  
The full expression of the latter is~\cite{Bini:2015xua}~\footnote{Since we neglect spin contributions beyond linear order, we do not include in $g_{S_*}^{\rm K}$ any dependence on the effective centrifugal radius $r_c$.}
\begin{align} 
g_{S_*}^K = \frac{1}{u} \Bigg\{
\sqrt{\frac{A^K}{W^K}}
\left[ 1 - 
\sqrt{\frac{A^K}{D^K}} \right]
- \frac{u\,  \partial_u A^K}{2  \left( 1 + \sqrt{W^K} \right)\sqrt{D^K}}
\Bigg\},
\label{gsstark}
\end{align}
where $A^{K}$ and $D^{K}$ are the EOB potentials in the limit $\nu\to0$ (with $D=\bar{D}^{-1}$) and
\begin{equation}
    W^K \equiv 1 + p_\varphi^2 u^2 + \frac{A^K}{D^K} p_r^2 .
\end{equation}

This gauge-fixing choice is motivated by the fact that, in the test-mass limit 
$\nu \to 0$, the function $g_S^{\rm DJS}$ trivially reduces to its Kerr value 
$g_S^{\rm K} = 2$, whereas $g_{S_*}^{\rm DJS}$ reproduces only the PN expansion 
of $g_{S_*}^{\rm K}$ in the circular limit.  
In contrast, in the $\overline{\rm DJS}$ spin gauge the test-mass limit is, by 
construction, enforced to reduce \emph{both} gyro-gravitomagnetic functions to 
their exact Kerr expressions for a spinning test particle.

It is important to stress that this gauge condition alone is not sufficient to 
fully fix the spin gauge.  
Since it constrains only the test-mass limit of the gyro-gravitomagnetic 
functions, it determines solely the $\mathcal{O}(\nu^{0})$ part of the 
spin–gauge coefficients.  
To completely fix the gauge, we therefore supplement the test-mass condition with 
the additional requirement that all remaining $\nu$-dependent components of the 
gauge coefficients vanish.\footnote{We also explored an alternative 
supplementary condition beyond the test-mass limit, namely the suppression of 
any dependence on $\dot{p}_{r}$.  
Since the resulting gyro-gravitomagnetic functions are numerically nearly 
indistinguishable from those obtained with our chosen condition, we adopt the 
latter for its greater simplicity.}

By imposing these conditions, we obtain a second system of equations that can be 
solved to fully determine all gauge coefficients; the corresponding solution is 
also provided in the supplementary material \cite{suppmat}.  
The resulting 5.5PN gyro-gravitomagnetic functions in the 
$\overline{\rm DJS}$ gauge are
\begin{widetext}
\begin{subequations}
\begin{align}
&\left(g_{S_*}^{5.5\text{PN}}\right)_{\overline{\rm DJS}}=
 \bigg( \frac{77}{256} + \frac{189}{64}\,\nu + \frac{1\,053}{256}\,\nu^2 + \frac{315}{64}\,\nu^3 + \frac{945}{256}\,\nu^4 \bigg) p_r^8 +  \bigg( \frac{145}{64} + \frac{1\,333}{64}\,\nu + \frac{4\,533}{256}\,\nu^2 + \frac{3\,759}{320}\,\nu^3 \cr
&+ \frac{16\,077}{1\,280}\,\nu^4 \bigg) p_r^6\,u  + \bigg[ \frac{77}{64} p_\varphi^2\,p_r^6 + \bigg( \frac{37}{16} + \frac{3\,297}{64}\,\nu + \frac{19\,467}{256}\,\nu^2 - \frac{11\,523}{64}\,\nu^3 - \frac{1\,377}{256}\,\nu^4 \bigg) p_r^4 \bigg] u^2 \cr
&+  \bigg[\frac{75}{16} p_\varphi^2\,p_r^4 + \bigg[-\frac{3}{16} + \bigg( \frac{28\,039}{80} - \frac{4\,829}{256} \pi^2 \bigg) \nu + \bigg( \frac{360\,199}{640} - \frac{123}{8} \pi^2 \bigg) \nu^2 - \frac{2\,931}{32} \nu^3 + \frac{171}{128} \nu^4 \bigg] p_r^2 \bigg] u^3 \cr
&+  \bigg[ -\frac{7}{8} + \frac{3}{8} p_\varphi^2\,p_r^2 + \frac{231}{128} p_\varphi^4\,p_r^4 + \bigg( \frac{9\,729}{1\,280} - \frac{123}{64} \pi^2 - \frac{3}{16} \mathsf{X}^{\nu^2}_{59} \bigg) \nu^2 - \frac{57}{64} \nu^3 - \frac{15}{256} \nu^4  + \nu \bigg( -\frac{7\,277\,321\,473}{11\,200}\cr
&+ \frac{1\,646\,928}{25} \gamma_{E} + \frac{26\,943}{2\,048} \pi^2 - \frac{5\,021\,084\,624}{189} \log 2 + \frac{1\,422\,585\,909}{140} \log 3 + \frac{3\,788\,359\,375}{756} \log 5 \bigg) \cr
& + \frac{823\,464}{25} \nu \log u \bigg] u^4 + \bigg\{ \frac{165}{64} p_\varphi^4\,p_r^2 + p_\varphi^2 \bigg[ \frac{5}{16} + \nu \bigg( \frac{2\,796\,846\,856}{525} - \frac{13\,503\,488}{25} \gamma_{E} + \frac{1\,023\,479\,286\,464}{4\,725} \log 2  \cr
&- \frac{2\,916\,358\,668}{35} \log 3 - \frac{7\,664\,875\,000}{189} \log 5 \bigg) - \frac{6\,751\,744}{25} \nu \log u \bigg] \bigg\} u^5  +  \bigg\{ \frac{77}{64} p_\varphi^6\,p_r^2 + p_\varphi^4 \bigg[ -\frac{3}{16}\cr
& + \nu \bigg( -\frac{57\,161\,452}{3} + 1\,930\,752\,\gamma_{E} - \frac{520\,306\,868\,128}{675} \log 2 + \frac{1\,490\,245\,128}{5} \log 3 + \frac{3\,870\,218\,750}{27} \log 5 \bigg)\cr
& + 965\,376 \nu \log u \bigg] \bigg) u^6  + \bigg[ \nu \bigg( \frac{996\,988\,856}{25} - \frac{100\,917\,248}{25} \gamma_{E} + \frac{43\,060\,589\,504}{27} \log 2 - \frac{3\,117\,148\,596}{5} \log 3 \cr
&- \frac{7\,891\,562\,500}{27} \log 5 \bigg)  - \frac{50\,458\,624}{25} \nu \log u \bigg] p_\varphi^{10}\,u^9  + \bigg[  \frac{5}{32} + \nu \bigg( \frac{582\,338\,384}{15} - 3\,932\,160\,\gamma_{E} \cr
&+ \frac{1\,055\,323\,161\,856}{675} \log 2 - \frac{3\,036\,137\,742}{5} \log 3 - \frac{7\,803\,406\,250}{27} \log 5 \bigg) - 1\,966\,080 \nu \log u \bigg] p_\varphi^6\,u^7\cr
&  + \bigg[  \nu \bigg( \frac{40\,346\,688}{7} - \frac{2\,916\,352}{5} \gamma_{E} + \frac{1\,085\,740\,722\,176}{4\,725} \log 2 - \frac{3\,152\,617\,362}{35} \log 3  - \frac{7\,929\,343\,750}{189} \log 5 \bigg)\cr
&- \frac{1\,458\,176}{5} \nu \log u \bigg] p_\varphi^{14} u^{11}  + \bigg[  \nu \bigg( -\frac{1\,507\,161\,628}{75} + \frac{50\,843\,648}{25} \gamma_{E} - \frac{541\,332\,696\,992}{675} \log 2 + 314\,079\,444 \log 3\cr
& + \frac{3\,958\,375\,000}{27} \log 5 \bigg) + \frac{25\,421\,824}{25} \nu \log u \bigg] p_\varphi^{12} u^{10}+ \bigg[  \frac{77}{256} + \nu \bigg( -\frac{739\,106\,572}{15} + 4\,988\,928\,\gamma_{E} \cr
&- \frac{266\,838\,974\,336}{135} \log 2 + \frac{1\,541\,014\,875}{2} \log 3  + \frac{19\,634\,453\,125}{54} \log 5 \bigg) + 2\,494\,464 \nu \log u \bigg] p_\varphi^8 u^8, \\
\cr
&\left(g_{S}^{5.5\text{PN}}\right)_{\overline{\rm DJS}}= p_r^4 u^2 \bigg( \frac{205}{4} \nu + \frac{30\,541}{128} \nu^2 - \frac{60\,813}{256} \nu^3 - \frac{107}{32} \nu^4 \bigg) + p_r^2 u^3 \bigg[ \bigg( \frac{1\,869\,859}{1\,600} - \frac{22\,301 \pi^2}{256} \bigg) \nu \cr&+ \bigg( \frac{877\,631}{960} - \frac{1\,087 \pi^2}{64} \bigg) \nu^2 - \frac{12\,579}{128} \nu^3 + \frac{3}{4} \nu^4 \bigg] + p_r^8 \bigg( - \frac{27}{64} \nu + \frac{99}{128} \nu^2 + \frac{945}{256} \nu^3 + \frac{189}{32} \nu^4 \bigg) \cr&+ p_r^6 u \bigg( - \frac{467}{64} \nu - \frac{1\,077}{128} \nu^2 - \frac{3\,607}{256} \nu^3 + \frac{1\,577}{160} \nu^4 \bigg) + p_\varphi^{10} u^9 \bigg[ \nu \bigg( \frac{4\,144\,026\,748}{75} - \frac{40\,997\,632}{5} \gamma_E\cr
& + \frac{304\,929\,978\,368 \log 2}{135} - \frac{22\,040\,831\,187 \log 3}{25} - \frac{11\,292\,265\,625 \log 5}{27} \bigg) - \frac{20\,498\,816}{5} \nu \log u \bigg]\cr
&+ p_\varphi^6 u^7 \bigg[ \nu \bigg( \frac{4\,033\,101\,256}{75} - 7\,987\,200 \gamma_E + \frac{1\,494\,627\,871\,744 \log 2}{675} - \frac{21\,466\,187\,217 \log 3}{25} \cr
&- \frac{11\,165\,609\,375 \log 5}{27} \bigg) - 3\,993\,600 \nu \log u \bigg] + p_\varphi^{14} u^{11} \bigg[ \nu \bigg( \frac{4\,193\,060\,296}{525} - 1\,184\,768 \gamma_E \cr
&+ \frac{1\,537\,720\,290\,304 \log 2}{4\,725} - \frac{22\,292\,460\,117 \log 3}{175} - \frac{11\,346\,546\,875 \log 5}{189} \bigg) - 592\,384 \nu \log u \bigg]\cr
&+ p_\varphi^2 u^5 \bigg[ \nu \bigg( \frac{258\,162\,588}{35} - \frac{5\,485\,792}{5} \gamma_E + \frac{1\,449\,515\,277\,856 \log 2}{4\,725} - \frac{20\,616\,856\,047 \log 3}{175} \cr
&- \frac{10\,966\,578\,125 \log 5}{189} \bigg) - \frac{2\,742\,896}{5} \nu \log u \bigg] + u^4 \bigg[ \bigg( \frac{313\,823}{5\,760} - \frac{1\,225 \pi^2}{384} - \frac{3}{16} \mathsf{X}^{\nu^2}_{59} \bigg) \nu^2 - \frac{239}{256} \nu^3 \cr
&- \frac{1}{32} \nu^4 + \nu \bigg( - \frac{90\,682\,596\,509}{100\,800} + \frac{2\,007\,488}{15} \gamma_E + \frac{62\,041 \pi^2}{1\,536} - \frac{35\,555\,719\,744 \log 2}{945} + \frac{20\,112\,254\,397 \log 3}{1\,400} \cr
&+ \frac{10\,839\,921\,875 \log 5}{1\,512} \bigg) + \frac{1\,003\,744}{15} \nu \log u \bigg] + p_\varphi^{16} u^{12} \bigg[ \nu \bigg( - \frac{524\,132\,537}{525} + 148\,096 \gamma_E\cr
&- \frac{192\,215\,036\,288 \log 2}{4\,725} + \frac{22\,292\,460\,117 \log 3}{1\,400}+ \frac{11\,346\,546\,875 \log 5}{1\,512} \bigg)+ 74\,048 \nu \log u \bigg]\cr
& + p_\varphi^4 u^6 \bigg[ \nu \bigg( - \frac{659\,682\,546}{25} + 3\,921\,840 \gamma_E - \frac{736\,894\,566\,512 \log 2}{675} + \frac{21\,071\,552\,787 \log 3}{50} \cr
&+ \frac{11\,075\,140\,625 \log 5}{54} \bigg) + 1\,960\,920 \nu \log u \bigg] + p_\varphi^{12} u^{10} \bigg[ \nu \bigg( - \frac{417\,671\,578}{15} + \frac{20\,655\,232}{5} \gamma_E\cr
&- \frac{766\,681\,745\,408 \log 2}{675} + \frac{22\,208\,583\,807 \log 3}{50} + \frac{11\,328\,453\,125 \log 5}{54} \bigg) + \frac{10\,327\,616}{5} \nu \log u \bigg] \cr
&+ p_\varphi^8 u^8 \bigg[ \nu \bigg( - \frac{1\,023\,922\,606}{15} + 10\,133\,760 \gamma_E - \frac{75\,583\,730\,432 \log 2}{27} + \frac{21\,791\,700\,297 \log 3}{20}\cr
&+ \frac{56\,189\,921\,875 \log 5}{108} \bigg) + 5\,066\,880 \nu \log u \bigg].
\end{align}
\end{subequations}
\end{widetext}
In this case, we have used the three dynamical variables $(u, p_r, p_\varphi)$ instead of $(u, p_r, p^2)$, as this choice simplifies the structure of our 
results.   The corresponding expressions in terms of $(u, p_r, p^{2})$ can be recovered 
straightforwardly by using the relation 
$p_\varphi = u \sqrt{p^{2} - p_{r}^{2}}$.

\subsection{Performance comparison: binding energy in the DJS and $\overline{\text{DJS}}$ spin gauges}

In this subsection, we assess the performance of the two spin gauges introduced 
above. To this end, we numerically compute the spin–orbit contribution to the 
binding energy, $E_b^{\rm SO}$, for circular orbits in both the DJS and 
$\overline{\rm DJS}$ gauges, and compare the resulting curves with those 
extracted from NR simulations.  
Although the circular-orbit binding energy is a gauge-invariant quantity and is, 
in principle, independent of the chosen spin gauge, this invariance holds 
strictly only up to the perturbative order to which the conservative dynamics is known.  
We therefore exploit the residual gauge dependence that arises when the binding 
energy is computed using truncated PN series, and use it as a diagnostic to 
compare the performance of the two spin gauges\footnote{Similar comparison 
strategies have been employed in 
Refs.~\cite{Khalil_2021,placidi}, and in particular in Ref.~\cite{Antonelli:2020aeb}.}.



To compute the analytical binding–energy curves from the EOB Hamiltonian, we 
begin by imposing the two circular-limit conditions
\begin{subequations}
    \begin{align}
        &\frac{d\hat{H}_{\rm EOB}(r,p_r,p_\varphi,\mathsf{X}^{\nu^2}_{59})}{dr}
        \bigg|_{p_r \to 0} = 0, \\
        &\left(
        \frac{d\hat{H}_{\rm EOB}(r,p_r,p_\varphi,\mathsf{X}^{\nu^2}_{59})}
             {dp_\varphi}
        \right)^{1/3}
        \bigg|_{p_r \to 0}
        = v_\omega ,
    \end{align}
\end{subequations}
which are solved simultaneously for $r$ and $p_\varphi$ for any given value of 
$v_\omega \equiv \Omega_\varphi^{1/3}$.  
The resulting values of $r$ and $p_\varphi$ are then substituted directly into 
the circular-orbit binding energy\footnote{We note that this differs from 
Eq.~\eqref{definition: binding energy}, where $\hat{H}_{\rm EOB}$, and thus 
$E_b$, is expressed in terms of $x$.}
\begin{equation}
    E_b(r,p_\varphi)
    = \hat{H}_{\rm EOB}\big|_{p_r \to 0}
      - \frac{1}{\nu},
\end{equation}
from which the spin–orbit contribution is isolated via the relation 
\cite{Ossokine_2018}
\begin{equation}
    E_b^{\rm SO}
        = \frac{1}{2}\big[\,E_b(\chi_1,\chi_2)
                         - E_b(-\chi_1,-\chi_2)\,\big],
    \label{spin-orbit binding nrgy}
\end{equation}
where we have made explicit the dependence on the two dimensionless black-hole 
spins, $\chi_1$ and $\chi_2$.

Once the EOB Hamiltonian and the values of the mass ratio and spins are 
specified, repeating the procedure described above for different values of the 
frequency parameter $v_\omega$ yields the linear-in-spin binding–energy curve 
$E_b^{\rm SO}(v_\omega)$.  
Fixing $\nu = 0.25$ and $\chi_1 = \chi_2 = 0.6$, we construct such curves in 
both spin gauges by systematically increasing the PN order of the spin–orbit 
sector of the Hamiltonian, starting from the leading 1.5PN contribution and 
progressively including higher-order terms up to 5.5PN.\footnote{For all PN 
orders below 5.5PN, we use the gauge-general expressions provided in 
Sec.~IIIC of Ref.~\cite{placidi}.}  
Since the linear-in-spin 5.5PN component of $\hat{H}_{\rm EOB}$ also depends on 
the unknown coefficient $\mathsf{X}^{\nu^2}_{59}$, we vary this parameter within 
the range $[-750,750]$, consistent with the $\mathcal{O}(10^{2})$ estimate 
reported in Ref.~\cite{Khalil_2021}.

Turning to the numerical binding-energy curve used as a reference, 
we rely on the data provided in Ref.~\cite{Ossokine_2018} (see Fig.~7 therein), 
where the NR curves are obtained with the Spectral Einstein Code from numerical 
simulations of the Simulating eXtreme Spacetimes catalog.

\begin{figure*}[t!]
    \centering
    \includegraphics[width=0.95\linewidth]{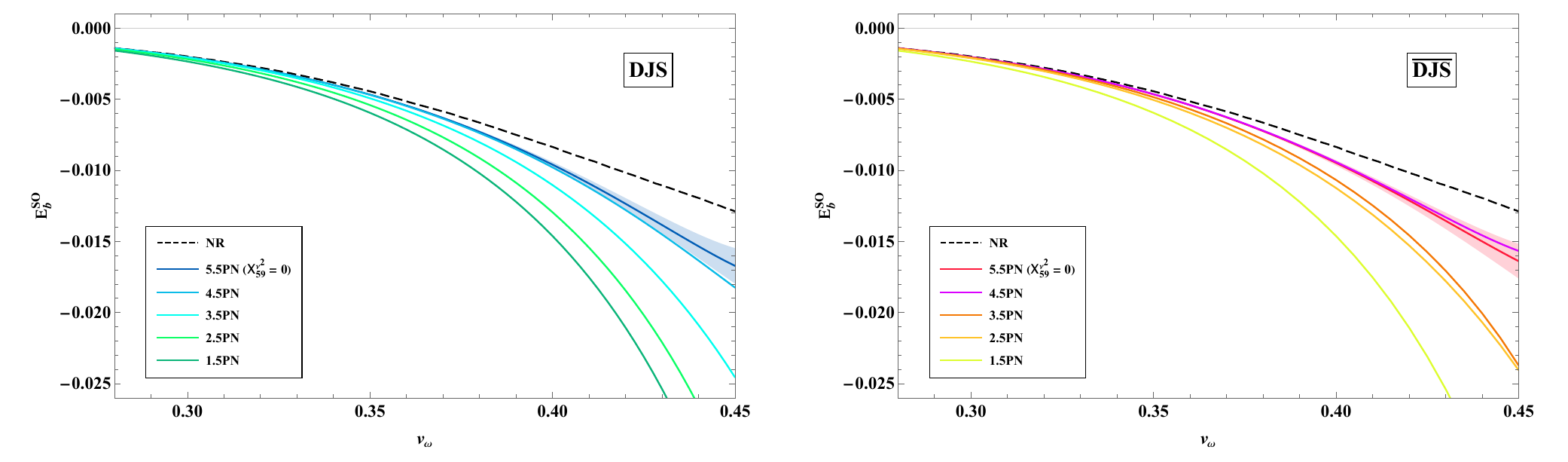}
   \caption{\label{fig:singlegauges}
Spin–orbit contribution to the binding energy for circular orbits as a function 
of $v_\omega$, shown in the DJS (left panel) and $\overline{\rm DJS}$ (right 
panel) spin gauges.  
In each panel, the numerical curve from Ref.~\cite{Ossokine_2018} is displayed as 
a black dashed line, together with the analytical curves obtained by truncating 
the spin–orbit sector of the EOB Hamiltonian at different PN orders (see the 
legend in the insets).  
At 5.5PN order, the $\mathsf{X}^{\nu^2}_{59}=0$ curve is accompanied by a shaded 
band indicating the variation induced by changing $\mathsf{X}^{\nu^2}_{59}$ 
within the range $[-750,750]$.  
The upper edge of the shaded region corresponds to 
$\mathsf{X}^{\nu^2}_{59}=750$, while decreasing $\mathsf{X}^{\nu^2}_{59}$ 
produces a systematic downward shift of the curves, with the lower edge reached 
at $\mathsf{X}^{\nu^2}_{59}=-750$.}
\end{figure*}

\begin{figure}[t!]
    \centering
    \includegraphics[width=0.91\linewidth]{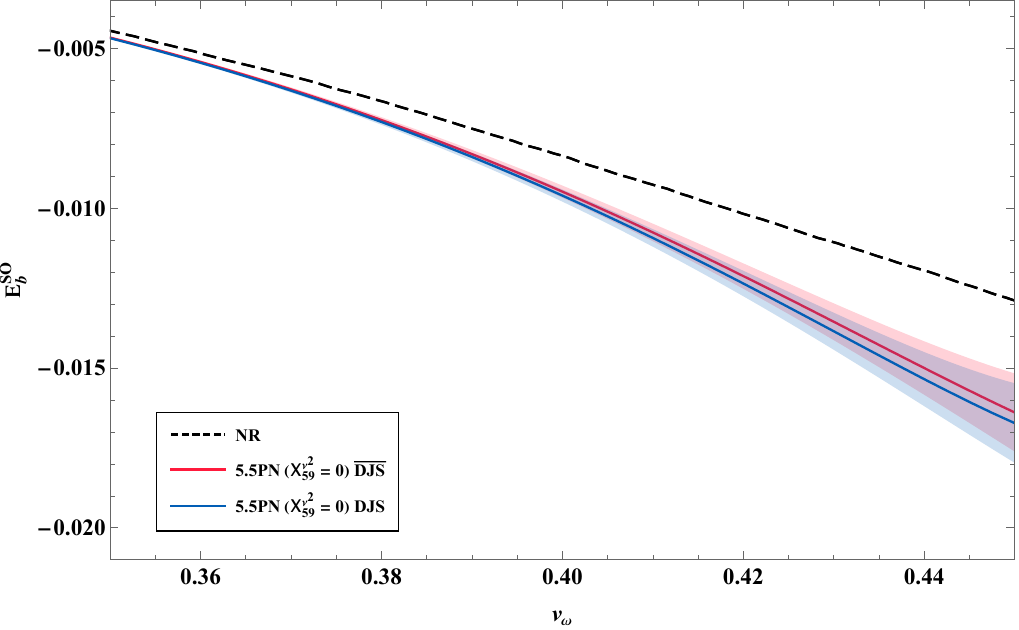}
   \caption{\label{fig:comparison}
Direct comparison between the 5.5PN curves shown in Fig.~\ref{fig:singlegauges} 
for the two spin gauges.  
To better emphasize the differences between the gauges, the range of 
$v_\omega$ displayed here is shorter than in Fig.~\ref{fig:singlegauges}.}
\end{figure}

The results of our analysis in both the DJS and $\overline{\rm DJS}$ gauges are 
presented in Figs.~\ref{fig:singlegauges} and \ref{fig:comparison}.

In Fig.~\ref{fig:singlegauges} we display, for both spin gauges, the binding 
energy obtained at different PN accuracies in the spin–orbit sector.  
We confirm that the DJS curves up to 4.5PN agree with those shown in Fig.~2 of 
Ref.~\cite{Antonelli:2020aeb}.  
Comparing the two gauges, we observe that the 1.5PN curves coincide, as expected, 
since spin gauge choices affect the dynamics only starting at next-to-leading 
order.  
From 2.5PN onward, each $\overline{\rm DJS}$ curve systematically lies above the 
corresponding DJS one, and is therefore slightly closer to the NR binding 
energy.

In Fig.~\ref{fig:comparison} we show a direct comparison between the 5.5PN 
curves obtained in the two spin gauges.  
We find that the $\overline{\rm DJS}$ gauge provides a slightly better agreement 
with the NR data across the entire explored range of 
$\mathsf{X}^{\nu^2}_{59}$ (represented by the shaded regions around the 
$\mathsf{X}^{\nu^2}_{59}=0$ curve), for both positive and negative values.

It is also interesting to note that the $\overline{\rm DJS}$ gauge exhibits a 
peculiar behavior at 4.5PN: its prediction is not only more accurate than the 
DJS curve at the same order, but even slightly more accurate than the 
$\overline{\rm DJS}$ 5.5PN curve with $\mathsf{X}^{\nu^2}_{59}=0$.  
Nevertheless, this 4.5PN curve lies consistently below the 5.5PN one when 
$\mathsf{X}^{\nu^2}_{59}=750$ (corresponding to the upper edge of the pink 
shaded region).  
This observation suggests that, if one expects the 5.5PN curve to systematically 
improve upon the previous order, the unknown parameter 
$\mathsf{X}^{\nu^2}_{59}$ should take a positive and sufficiently large value.

\section{Conclusions}
\label{sec:concl}

In this work, we have derived the gauge-general expressions for the two 
gyro-gravitomagnetic functions, $g_S$ and $g_{S_*}$, entering the spin–orbit 
sector of the EOB Hamiltonian, up to 5.5PN order.  
Our derivation includes both local and nonlocal-in-time contributions, thereby 
providing a complete characterization of the spin–orbit coupling at this level 
of accuracy.  
These results extend those of previous works 
\cite{Antonelli_2020,Khalil_2021,placidi} either by increasing the PN accuracy 
or by lifting the gauge restrictions adopted therein.

We then used these expressions to compute two gauge-invariant quantities for 
quasi-circular orbits: the binding energy and the fractional periastron advance.  
This constitutes a nontrivial consistency check of our gauge-general formulas, 
as it confirms that all dependence on the gauge coefficients cancels out when 
constructing physical observables.

Finally, we used our results to compare the performance of the DJS and 
$\overline{\rm DJS}$ spin gauges in the specific case of an equal-mass binary 
with spins $\chi_1 = \chi_2 = 0.6$.  
This analysis is shown in Figs.~\ref{fig:singlegauges} and 
\ref{fig:comparison}, where we plot the spin–orbit contribution to the 
circular-orbit binding energy, $E_b^{\rm SO}$, as a function of the frequency 
parameter $v_\omega$.  
By comparing the analytical curves at various PN orders in the spin–orbit 
sector with NR data, we find that, starting at 2.5PN order, each 
$\overline{\rm DJS}$ curve systematically lies slightly above the corresponding 
DJS curve, thereby providing a closer agreement with the NR results.  
Furthermore, the fact that, within the $\overline{\rm DJS}$ gauge, the 5.5PN 
curve improves upon the 4.5PN one only when the undetermined coefficient 
$\mathsf{X}^{\nu^2}_{59}$ takes positive and sufficiently large values may be 
interpreted as an indication of the expected magnitude and sign of this 
coefficient.

The main outcome of our, albeit non-conclusive, exploration is that the 
$\overline{\rm DJS}$ spin gauge may offer advantages for the description of 
quasi-circular inspirals within EOB-based waveform models.  
A natural direction for future work is to investigate whether the trends 
observed here, derived for a specific binary configuration, persist across binaries with different mass ratios and with 
spins of varying magnitudes and orientations.

\section*{Acknowledgments}
We acknowledge Marta Orselli, Marta Cocco, Elisa Grilli and Davide Panella for many valuable discussions. G. Grignani and A. Placidi acknowledge financial support from the Italian Ministry of University and Research (MUR) through the program ``Dipartimenti di Eccellenza 2023-2027" (Grant SUPER-C), from ``Fondo di Ricerca d'Ateneo"  2023 (GraMB) of the University of Perugia and from the Italian Ministry of University and Research (MUR) via the PRIN 2022ZHYFA2, GRavitational wavEform models for coalescing compAct binaries with eccenTricity (GREAT).  L. Sebastiani thanks Alex Segaricci and Leonardo Dinoi for their insightful and inspiring comments during the development of this work.
L. Sebastiani's research is supported by the European Research Council (ERC) Horizon Synergy Grant “Making Sense of the Unexpected in the Gravitational-Wave Sky” grant agreement no. GWSky–101167314.

\appendix

\section{Quasi-Keplerian parametrization}\label{app.qk}

The quasi-Keplerian parametrization was first introduced by Damour and Deruelle 
in Ref.~\cite{qk_DD_AIHPA_1985__43_1_107_0}, with spin contributions later 
included in Refs.~\cite{qkspin1_Tessmer_2010, qkspin2_Tessmer_2012}.  
Here, we review this parametrization at leading non-spinning and linear-in-spin 
order, which is sufficient for the computation in Sec.~\ref{sec: non loc}.  
For completeness, we also refer the reader to Ref.~\cite{qk2_Memmesheimer_2004} 
for its 3PN extension in the non-spinning case.

This parametrization is constructed to reduce, at Newtonian order, to the 
classical Keplerian parametrization \cite{brouwer1961methods}, namely
\begin{subequations}
\begin{align}
r &= a_r(1 - e \cos u_e), \label{faker}\\[6pt]
\varphi - \varphi_0 &= v \equiv 2 \arctan \left[ \left( \frac{1 + e}{1 - e} \right)^{1/2} 
\tan \frac{u_e}{2} \right], \label{fakephi}\\[6pt]
\ell &\equiv n(t - t_0) = u_e - e \sin u_e, \label{fakekepler}
\end{align}
\end{subequations}
where $r$ and $\varphi$ are the usual polar coordinates in the orbital plane, 
with $\mathbf{r} = r(\cos\varphi, \sin\varphi, 0)$.  
Here, $a_r$ denotes the semi-major axis and $e$ is the orbital eccentricity.  
The auxiliary variables $u_e$ and $v$ are the \emph{eccentric anomaly} and 
\emph{true anomaly}, respectively.  
The time dependence is encoded in Kepler’s equation \eqref{fakekepler}, where 
$\ell$ is the \emph{mean anomaly} and $n = 2\pi/T$ is the \emph{mean motion}, 
with $T$ the orbital period.  
The constants $t_0$ and $\varphi_0$ give the initial values of the corresponding 
variables and can be set to zero by an appropriate choice of reference frame.

The quantities $a_r$, $e$, and $n$ are directly related to gauge-invariant 
dynamical quantities—namely the $\mu$-reduced non-relativistic energy 
$\bar{E} = E - 1/\nu$ and the reduced angular momentum $L$—through
\begin{equation}
a_r = \frac{1}{-2\bar{E}}\, , 
\qquad
e = \sqrt{1 + 2 \bar{E} L^{2}}\, , 
\qquad
n = (-2\bar{E})^{3/2}\, .
\end{equation}
When generalizing this parametrization to include the leading spin–orbit 
coupling, the above relations become \cite{qkspin2_Tessmer_2012}
\begin{subequations}\label{quasiKeplerian}
\begin{align}
r &= a_r(1 - e_r \cos u_e), \label{appendice:rqk} \\[4pt]
\varphi &= 2K \arctan \left[ \sqrt{\frac{1 + e_\varphi}{1 - e_\varphi}} 
\tan \frac{u_e}{2} \right], \\[4pt]
\ell &\equiv nt = u_e - e_t \sin u_e,
\end{align}
\end{subequations}
where we have set $t_0 = \varphi_0 = 0$.  
We now have to distinguish between the three eccentricities 
$(e_r, e_\varphi, e_t)$, i.e.\ the radial, angular, and time eccentricity, 
respectively.  
The factor $K$ denotes the fractional periastron advance per radial period and 
encodes the precession of the orbit.


The relation between the orbital parameters and the gauge-invariant quantities 
can be derived from the rescaled Hamiltonian at leading non-spinning and 
linear-in-spin order,
\begin{equation} \label{eq:Hlo}
    \bar{H} = \frac{p^2}{2} - \frac{1}{r} 
    + \frac{L}{c^{3} r^{3}} \left[ 2\delta \chi_A - (\nu - 2)\chi_S \right],
\end{equation}
where $\bar{H} = H - c^2/\nu$ denotes the dimensionless non-relativistic Hamiltonian.  
By inserting the quasi-Keplerian parametrization \eqref{quasiKeplerian} into 
Eq.~\eqref{eq:Hlo} and evaluating it at periastron ($u_e = 0$) and apastron 
($u_e = \pi$), one obtains the relations
\begin{align}
\bar{E} &= -\frac{1}{2a_r}
- \frac{(2-\nu)\chi_S + 2\delta\chi_A}{2\, c^3\, a_r^{5/2}\!\sqrt{1 - e_r^{2}}}, 
\label{app:ebat} \\[6pt]
L &= \sqrt{a_r(1 - e_r^{2})}
- \frac{(e_r^{2} + 3)\left[\,2\delta\chi_A + (2 - \nu)\chi_S\,\right]}
       {2\, c^3\, a_r(1-e_r^{2})}. 
\label{app:mom_ang}
\end{align}

Inverting these equations for $e_r$ and $a_r$ yields
\begin{align}
e_r &= \sqrt{1 + 2 \bar{E} L^2} 
+ \frac{4 \bar{E} (1 + \bar{E} L^2)}{\, c^3L \sqrt{1 + 2 \bar{E} L^2}} 
\left[ 2 \delta \chi_A + (2 - \nu) \chi_S \right], \label{app.er} \\[6pt]
a_r &= \frac{-1}{2 \bar{E}} + \frac{2 \delta \chi_A + (2 - \nu) \chi_S}{\, c^3 L}. \label{app:ar}
\end{align}

For the mean motion and fractional periastron advance, starting from the definitions
\begin{align}
T_r &= \oint \frac{dr}{\dot{r}} 
= 2  \int_{r(u_e=0)}^{r(u_e=\pi)} \frac{dr}{\partial H / \partial p_r}, \\[8pt]
K &= \frac{1}{2\pi} \oint \frac{dr}{\dot{r}} \dot{\varphi}
= 2 \int_{r(u_e=0)}^{r(u_e=\pi)} dr \, \frac{\partial H / \partial L}{\partial H / \partial p_r},
\end{align}
we obtain
\begin{align}
n &= \frac{2\pi}{T_r} = 2\sqrt{2}(-\bar{E})^{3/2} \\
  &= \frac{1}{a_r^{3/2}} + \frac{3\left[ 2\delta \chi_A + (2 - \nu)\chi_S \right]}{2 \,c^3 a_r^{3} \sqrt{1 - e_t^2}}, \label{app:n} \\[10pt]
K &= 1 - \frac{2(2-\nu)\chi_S + 4\delta \chi_A}{\, c^3 L^3} \\
  &= 1 - \frac{4\delta \chi_A + 2(2 - \nu)\chi_S}{c^3\,a_r^{3/2} (1 - e_t^2)^{3/2}}. \label{app:k}
\end{align}

Finally, the three eccentricities are related by
\begin{subequations}
\begin{align}
\frac{e_r}{e_t} &= 1 + \frac{2\bar{E}}{c^3 \,L} 
\left[ 2\delta \chi_A + (2 - \nu)\chi_S \right], 
\label{app:er/et} \\[6pt]
\frac{e_\varphi}{e_t} &= 1 + \frac{2\bar{E}}{c^3 \,L} 
\left[ 2\delta \chi_A + (2 - \nu)\chi_S \right], 
\label{app:ephi/et}
\end{align}
\end{subequations}
from which it is evident that $e_r$, $e_\varphi$, and $e_t$ all coincide at 
leading Newtonian order.  
If we further substitute into 
Eqs.~\eqref{app:er/et}–\eqref{app:ephi/et} the expressions for the energy and 
angular momentum given in Eqs.~\eqref{app:ebat}–\eqref{app:mom_ang}, and solve 
iteratively for $e_t$, we obtain
\begin{equation}
e_r = e_\varphi
= e_t 
+ \frac{e_t\!\left[(\nu - 2)\chi_S - 2\delta\chi_A\right]}
       {c^3\, a_r \sqrt{a_r(1 - e_t^2)}}
\, .
\label{all_in_et}
\end{equation}

\newpage
\bibliography{Bibliography.bib,local.bib,refs.bib} 

\end{document}